\documentclass[fleqn,usenatbib]{mnras}

\usepackage[T1]{fontenc}
\usepackage{ae,aecompl}

\usepackage{graphicx}	% Including figure files
\usepackage{amsmath}	% Advanced maths commands
\usepackage{amssymb}	% Extra maths symbols

\usepackage{txfonts} %old version of newtxtext, should be loaded after amsmath

\usepackage{epstopdf}
\usepackage{journals}

\usepackage[none]{hyphenat}
\title[Two-component gravitational instability in spiral galaxies]{Two-component gravitational instability in spiral galaxies}
\author[A. A. Marchuk and N. Y. Sotnikova]
{A. A. Marchuk$^{1}$\thanks{E-mail:
a.marchuk@spbu.ru (AAM)} and N. Y. Sotnikova$^{1,2}$\\
$^{1}$St. Petersburg State University,
Universitetskij pr.~28, 198504 St. Petersburg, Stary Peterhof, Russia\\
$^{2}$Isaac Newton Institute of Chile, St. Petersburg Branch, Russia\\
}

\date{Accepted XXX. Received YYY; in original form ZZZ}

\pubyear{2016}

\begin{document}
\label{firstpage}
\pagerange{\pageref{firstpage}--\pageref{lastpage}}
\maketitle

%%%%%%%%%%%%%%%%%%%%%%%%%%%%%%%%%%%%%%%%%%%%%%%%%%%%%%%%%%%%%%%%%%%%%%
\begin{abstract}
We applied a criterion of gravitational instability, valid for two-component and  infinitesimally thin discs, to observational data along the major axis for 7 spiral galaxies of early types. Unlike most papers, the dispersion equation corresponding to the criterion was solved directly without using any approximation. The velocity dispersion of  stars in the radial direction $\sigma_R$ was limited by the range of possible values instead of a fixed value. For all galaxies, the outer regions of the disc were analyzed up to 
$R \le 130 \arcsec$. The maximal and sub-maximal disc models were used to translate surface brightness into surface density. The largest destabilizing disturbance stars can exert on a gaseous disc was estimated. It was shown that the two-component criterion differs a little from the one-fluid criterion for galaxies with a large surface gas density, but it allows to explain large-scale star formation in those regions where the gaseous disc is stable. In the galaxy NGC~1167 star formation is entirely driven by the self-gravity of the stars. A comparison is made with the conventional approximations which also include the thickness effect and with models for different sound speed $c_\mathrm{g}$. It is shown that values of the effective Toomre parameter correspond to the instability criterion of a two-component disc $Q_\mathrm{eff}<1.5-2.5$. This result is consistent with previous theoretical and observational studies.
\end{abstract}
%%%%%%%%%%%%%%%%%%%%%%%%%%%%%%%%%%%%%%%%%%%%%%%%%%%%%%%%%%%%%%%%%%%%%

%%%%%%%%%%%%%%%%%%%%%%%%%%%%%%%%%%%%%%%%%%%%%%%%%%%%%%%%%%%%%%%%%%%%%
%%%%%%%%%%%%%%%%%%%%%%%%%%%%%%%%%%%%%%%%%%%%%%%%%%%%%%%%%%%%%%%%%%%%%

%%%%%%%%%%%%%%%%%%%%%%%%%%%%%%%%%%%%%%%%%%%%%%%%%%%%%%%%%%%%%%%%%%%%%
\section{Introduction}
%%%%%%%%%%%%%%%%%%%%%%%%%%%%%%%%%%%%%%%%%%%%%%%%%%%%%%%%%%%%%%%%%%%%%

Several mechanisms that can explain the nonlinear relation between star formation rate (SFR) and surface gas density $\Sigma_\mathrm{g}$ in disc galaxies (the so-called Schmidt law, \citealp{Schmidt1959}) are known. Among these mechanisms, the following are often discussed: gravitational instability, destruction of giant molecular clouds (GMC) by differential rotation (galactic shear), thermal instability and subsequent molecularization of the resulting cold clouds (see, for example, \citealt{Leroy_etal2008}). All these mechanisms are threshold processes what means that star formation starts when gas is cold and dense enough.

Thorough analysis conducted by \citet{Leroy_etal2008} did not reveal the leading process, which would be able to explain large-scale star formation in galaxies completely. However, as it was shown for the first time in \citet{Kennicutt1989},  gravitational instability itself gives a good agreement with an observational data for large number of galaxies. In many subsequent investigations this result was confirmed. Recent investigations have shown that the link between SFR and gravitational instability is most likely indirect. Many other processes are involved in the connection (turbulence, transporting of mass and angular momentum, stellar feedback, external accretion) which all together lead to a roughly constant depletion time in the molecular gas, the main building material for stars (e.g., \citealp{Kennicutt_Evans2012,Forbes_etal2014,Goldbaum_etal2015,Goldbaum_etal2016}). However the direct effect of the gravitational instability can be rather easily calculated and unstable regions can be compared with star formation areas.

The connection between Schmidt law and one-fluid criterion of gravitational instability was investigated for a large sample of galaxies in \citet{Kennicutt1989}. As was originally suggested by \citet{Quirk1972}, if the origin of the empirical Schmidt law is related to gravitational instability, it is expected to break down when the surface density of the gas falls below the critical surface density. \citet{Kennicutt1989} used the intensity of the H$\alpha $ line as the indicator of star formation. For regions with large amount of gas he obtained a nonlinear relationship $I(\mathrm{H}\alpha) \propto \Sigma_\mathrm{g}^{1.3}$. Noticeable star formation is observed in exactly the same region where the surface gas density $\Sigma_\mathrm{g}$ lies above the critical value $\Sigma_\mathrm{g}^\mathrm{cr89} = 
\displaystyle \alpha \frac{\varkappa c_\mathrm{g}}{\pi G}$, where $\varkappa$ is epicyclic frequency and $c_\mathrm{g} $ is the sound speed in the gas. If $\alpha = 1$ this criterion becomes identical to one predicted by the one-fluid gravitational instability model  with respect to axisymmetric perturbations
$\Sigma_\mathrm{g} \ga 
\displaystyle \Sigma_\mathrm{g}^\mathrm{cr} \equiv \frac{\varkappa c_\mathrm{g}}{\pi G}$ which also can be written in terms of the dimensionless parameter $Q_\mathrm{g} \equiv 
\displaystyle \frac{\varkappa c_\mathrm{g}}{\pi G \Sigma_\mathrm{g}} \la 1$ \citep{G_LB1965}.

\citet{Kennicutt1989} empirically obtains the value of coefficient $\alpha \approx 0.5$\footnote{\citet{Kennicutt1989} quoted a value $\alpha = 0.67$, but used a constant of 3.36 rather than $\pi$ in the definition of $\Sigma_\mathrm{g}^\mathrm{cr89}$, and the one-dimensional velocity dispersion $\sigma_\mathrm{g}$ instead of $c_\mathrm{g} = \sigma_\mathrm{g} \gamma^{1/2}$, with $\gamma = 5/3$ the adiabatic index. The later is correct only for the neutral gas with the velocity dispersion that is not dominated by turbulence. With the above correction $\alpha = 0.5$ 
\citep{Schaye2004}.} and emphasizes that in general, one expects $\alpha < 1$ in a realistic gas/stellar disc, due to the onset of two fluid instabilities \citep{Jog_Solomon1984}. In fact, classical criterion with $\alpha = 1$ \citep{G_LB1965} is applicable only for axisymmetric perturbations in the infinitely thin gas or stellar disc. Since there are perturbations of higher modes (nonaxisymmetric) in the disc it requires a higher level of stability in order to remain stable. Theoretical analysis (see, e.g., \citealp{Morozov1985,Polyachenko_etal1997}) 
and numerical experiments 
(see, e.g. \citealp{Khoperskov_etal2003,Li_etal2005}) confirm this statement. Therefore empirically obtained factor $\alpha\approx 0.5 $ ($Q_\mathrm{g} \le 2 $) corrects the \citet{G_LB1965} criterion for nonaxisymmetric perturbations.

In subsequent works the threshold relationship between SFR and gravitational instability of gaseous disc was repeatedly investigated and confirmed (for example, \citealp{Martin_Kennicutt2001,Hunter_etal1998} found $\alpha = 0.37-0.53 $). According to this approach it is possible to explain star formation in many galaxies containing a large amount of gas. However a galaxy-to-galaxy scatter in $\alpha$ at the threshold region is found to be relatively large. \citet{Martin_Kennicutt2001,Hunter_etal1998} suggested that $\alpha$ may vary systematically along the Hubble sequence. More severe problem is that a number of disc galaxies have discs that are subcritical i.e. have 
$\Sigma_\mathrm{g} < \Sigma_\mathrm{g}^\mathrm{cr89}$ over their entire extent, yet reveal widespread star formation.

\citet{Jog_Solomon1984} introduced two-fluid gravitational instability model. Gaseous disc was considered together with stellar disc and the whole problem was solved in hydrodynamic approximation. Accurate gravitational instability criterion for gas in the presence of collisionless stellar disc (two-component criterion) was obtained by \citet{Rafikov2001}. It can be formulated in terms of dimensionless parameter $Q_\mathrm{eff}$ which is represented as nonlinear function of $Q_\mathrm{g}$ and $Q_\mathrm{s} \equiv \displaystyle \frac{\varkappa \sigma_R}{3.36 G \Sigma_\mathrm{s}}$ \citep{Toomre1964}, where $\sigma_R $ is the stellar velocity dispersion in the radial direction and $\Sigma_\mathrm{s}$ is the stellar surface density. The main result obtained by \citet{Jog_Solomon1984, Rafikov2001} is that the presence of a stellar disc changes the dynamic state of the gas, making it less stable even when $Q_\mathrm{s} > 1$. Hence the two-component criterion of gravitational instability can be applied to those galaxies in which star formation is noticeable, while one-component criterion gives a stable gas disc.

\citet{Boissier_etal2003} determined radial profiles of  $Q_\mathrm{eff}$ for 16 galaxies with known rotation curves using two-fluid instability criterion in the approximate form given by \citet{Wang_Silk1994}. In all cases these profiles were above those obtained according to the one-fluid instability criterion.

\citet{Leroy_etal2008} showed that $Q_\mathrm{eff}$ exhibits a much narrower range of values than $Q_\mathrm{g}$ and that in the presence of stars gas becomes only marginally stable with $Q_\mathrm{eff} = 1.3-2.5$. Two-fluid gravitational instability criterion has been applied to individual galaxies in many works \citep{Romeo_Wiegert2011,Meurer_etal2013,Zheng_etal2013,Westfall_etal2014,Yim_etal2014,Tenjes_etal2017}. The exact two-component criterion \citep {Rafikov2001} and various approximations of the two-fluid criterion \citep{Wang_Silk1994,Romeo_Wiegert2011} such as multicomponent approximation with molecular gas \citep{Romeo_Falstad2013, Romeo_Fathi2016, Hallenbeck_etal2016, Romeo_Mogotsi2017} were used. Various types of galaxies like dwarfs \citep{Elson_etal2012}, LSB \citep{Garg_Banerjee2017} and very bright galaxies \citep{Hunter_etal2013} were considered in stability analysis. In a number of papers \citep{Zheng_etal2013, Yim_etal2014, Romeo_Fathi2016, Romeo_Mogotsi2017} gas velocity dispersion profiles were taken into account.

There are several difficulties in applying the two-component instability criterion to observed galaxies. For example, it is  difficult to take into account the thickness of the stellar disc. 
\citet{Romeo_Wiegert2011} showed that when the finite disc thickness is taken into account, the effective parameter $Q_\mathrm{eff}$ is 20-50\% greater than in the case of the infinitely thin disc. On the other hand, \citet{Elmegreen2011} analyzed the effect of gas dissipation on the disc stability in the gas-stars case and found that the dissipation effect partially compensates the stabilizing effect of the disc thickness, leading to the same threshold value of  $Q_\mathrm{eff}\approx $ 2-3 as in the case of nonaxisymmetric perturbations.

Another factor which can lead to overestimated values of $Q_\mathrm{eff}$ is azimuthal data averaging. In galaxies hydrogen distribution is often clumpy and the filling factor can vary from 6 to 50 \% \citep{Braun1997}. \citet{Martin_Kennicutt2001} have demonstrated that using azimuthally averaged gas densities and SFRs for one-fluid criterion can lead to errors in $\alpha$ near the threshold radius as large as a factor of 2 when the disc is highly nonaxisymmetric. This conclusion is also true for gas-stars instability. \citet{Yang_etal2007} studied Large Magellanic Cloud (LMC) and obtained $Q_\mathrm{eff}$ maps using exact gravitational criterion \citep{Rafikov2001} for constant velocity dispersion in radial direction $\sigma_R = 15$~km\,s$^{-1}$. These maps show that young stellar clusters always lie inside regions constrained by contours $Q_\mathrm {eff} < 1.0$.

As a rule, when the criterion of the gas-stars instability is applied to specific galaxies, approximation \citet{Wang_Silk1994} or \citet{Romeo_Wiegert2011, Romeo_Falstad2013} is used rather than exact formula. In addition $\sigma_R$ profile is often determined not from spectral data, but from various empirical relationships and such approach leads to errors. There are only a few works where the gas-stars criterion was applied with stellar velocity dispersion profiles $\sigma_R$ 
recovered from the line-of-sight kinematics \citep{Silchenko_etal2011, Romeo_Fathi2016}. There are also uncertainties in conversion of surface stellar brightness in stellar surface density.

The motivation for this article is to apply the two-component gravitational instability criterion in its most accurate form (in the kinetic approximation \citealp{Rafikov2001}) to a number of galaxies, taking into account the complete data on gas and stars including the observed profiles of the velocity dispersion of stars and the surface densities of stellar discs determined from rotation curves.

In Section 2 we give two-component instability theoretical background in general. In Section 3 we list the galaxies from the sample and provide the data used. In Section 4 we describe data reduction and $Q_\mathrm{eff}$ profiles calculation. In Section 5 we present our main results and discuss the uncertainties of our method. In Section 6 we give some conclusions about the instability level. In Appendix we describe in details the galaxies from the sample.

%%%%%%%%%%%%%%%%%%%%%%%%%%%%%%%%%%%%%%%%%%%%%%%%%%%%%%%%%%%%%%%%%%%%%
\section{Two-component instability. Theory}
%%%%%%%%%%%%%%%%%%%%%%%%%%%%%%%%%%%%%%%%%%%%%%%%%%%%%%%%%%%%%%%%%%%%%

%%%%%%%%%%%%%%%%%%%%%%%%%%%%%%%%%%%%%%%%%%%%%%%%%%%%%%%%%%%%%%%%%%%%%
\subsection{Hydrodynamic approximation}
%%%%%%%%%%%%%%%%%%%%%%%%%%%%%%%%%%%%%%%%%%%%%%%%%%%%%%%%%%%%%%%%%%%%%

\citet{Jog_Solomon1984} solved the hydrodynamic equations describing an azimuthally symmetric galactic disc as a two-fluid system. The criterion of the disc instability was formulated as follows\footnote{\cite{Rafikov2001} has written the criterion in this useful form.}
%%%%%%%%%%%%%%%%%%%%%%%%%%%%%%%%%%%%%%%%%%%%%%%%%%%%%%%%%%%%%%%%%%%%%
\begin{equation}
\frac{1}{Q(\bar{k})} \equiv 
\frac{2}{Q_\mathrm{s}}\frac{\bar{k}}{1 + \bar{k}^2} + 
\frac{2}{Q_\mathrm{g}}s\frac{\bar{k}}{1 + \bar{k}^2 s^2} > 1 \, ,
\label{eq:2fluidJS1}
\end{equation}
%%%%%%%%%%%%%%%%%%%%%%%%%%%%%%%%%%%%%%%%%%%%%%%%%%%%%%%%%%%%%%%%%%%%%
for at least some dimensionless wave numbers $\bar{k} \equiv k \sigma_\mathrm{s} / \varkappa$. In the above expression \eqref{eq:2fluidJS1} $\sigma_\mathrm{s}$ represents the stellar velocity dispersion\footnote{In the hydrodynamic approximation, the velocity distribution is assumed to be isotropic, but we mean by $\sigma_\mathrm{s}$ the velocity dispersion of stars in the radial direction.}, ${\displaystyle Q_\mathrm{s} \equiv 
\frac{\varkappa\,\sigma_\mathrm{s}}{\pi\, G\,\Sigma_\mathrm{s}}}$ is the dimensionless Toomre parameter for a stellar disc\footnote{
Introduced by analogy with $Q_\mathrm{g}$, the parameter $Q_\mathrm{s}$ differs from the value given for collisionless systems \citep{Toomre1964}. This means that, with the notation introduced, the one-component stellar disc is unstable with respect to axisymmetric perturbations, when 
$Q_\mathrm{s} < 3.36/\pi = 1.07$.}, 
and $s \equiv c_\mathrm{g} / \sigma_\mathrm{s}$ is 
the ratio of gas sound speed to stellar velocity dispersion. The maximum of $Q^{-1}(\bar{k})$ in the expression~\eqref{eq:2fluidJS1} can be denoted as $Q^{-1}_\mathrm{eff}$. Thus the criterion of the disc instability with respect to axisymmetric perturbations can be expressed as $Q_\mathrm{eff} < 1$.

The main result obtained by \citet{Jog_Solomon1984} is that under certain conditions even for a stable gaseous disc ($Q_\mathrm{g}>1$) and a stable stellar disc ($Q_\mathrm{s}>1$), the combined stellar-gas disc can be  unstable ($Q_\mathrm{eff} <1 $).

\citet{Elmegreen1995} has shown that the equality $Q_\mathrm{eff} = 1$ may be reduced to the cubic equation for the unknown $\bar{k}$. \citet{Efstathiou2000} rewrites the criterion~\eqref{eq:2fluidJS1} in terms of the gas critical surface density $\Sigma_\mathrm{g}^\mathrm{cr,2}$ 
%%%%%%%%%%%%%%%%%%%%%%%%%%%%%%%%%%%%%%%%%%%%%%%%%%%%%%%%%%%%%%%%%%%%%
\begin{equation}
\Sigma_\mathrm{g}^\mathrm{cr,2} = 
\frac{\varkappa c_\mathrm{g}}{\pi\, G\, g(a,b)} = 
\frac{\Sigma_\mathrm{g}^\mathrm{cr}}{g(a,b)} \, ,
\label{eq:2fluid}
\end{equation}
%%%%%%%%%%%%%%%%%%%%%%%%%%%%%%%%%%%%%%%%%%%%%%%%%%%%%%%%%%%%%%%%%%%%%
where
$a = \sigma_\mathrm{s} / c_\mathrm{g} = 1 / s$, 
$b = \Sigma_\mathrm{s} / \Sigma_\mathrm{g}$,
$g(a,b)$ is a numerically calculated function. 
Since $g(a, b)>1$ for any $a$ and $b$ therefore the instability in the presence of a stellar disc comes for smaller surface gas densities than in the one-fluid model.

\citet{Wang_Silk1994} have proposed an approximation formula for the effective parameter $Q_\mathrm{eff}$ and expressed it  through $Q_\mathrm{s}$ and $Q_\mathrm{g}$
%%%%%%%%%%%%%%%%%%%%%%%%%%%%%%%%%%%%%%%%%%%%%%%%%%%%%%%%%%%%%%%%%%%%%
\begin{equation}
\frac{1}{Q_\mathrm{eff,WS}} = 
\frac{1}{Q_\mathrm{s}} + \frac{1}{Q_\mathrm{g}} \,.
\label{eq:WS}
\end{equation}
%%%%%%%%%%%%%%%%%%%%%%%%%%%%%%%%%%%%%%%%%%%%%%%%%%%%%%%%%%%%%%%%%%%%%
\citet{Romeo_Wiegert2011} showed that this formula can lead to the error in $Q_\mathrm{eff}$ up to 50 \%. They derived and justified their own approximation formula
%%%%%%%%%%%%%%%%%%%%%%%%%%%%%%%%%%%%%%%%%%%%%%%%%%%%%%%%%%%%%%%%%%%%%
\begin{equation}
    \frac{1}{Q_\mathrm{eff, RW1}}=
    \begin{cases}
    \displaystyle
      \frac{W}{Q_\mathrm{s}} + \frac{1}{Q_\mathrm{g}},\: 
      Q_\mathrm{s} \ge Q_\mathrm{g} \, , \\[1.0em]
    \displaystyle
      \frac{1}{Q_\mathrm{s}} + \frac{W}{Q_\mathrm{g}},\: 
      Q_\mathrm{g} \ge Q_\mathrm{s} \, ,
    \end{cases}
\label{eq:RW1}
\end{equation}
%%%%%%%%%%%%%%%%%%%%%%%%%%%%%%%%%%%%%%%%%%%%%%%%%%%%%%%%%%%%%%%%%%%%%
where $W=\displaystyle\frac{2s}{1+s^2}$ is a dimensionless weight function. In both approximations \citep{Wang_Silk1994,Romeo_Wiegert2011} 
$Q_\mathrm{s}$ is calculated for the  factor $\pi$ instead of 3.36.

All these approximations are not very essential since modern computational tools allow to easily find $Q_\mathrm{eff}$ directly from the dispersion equation for given parameters. In addition, criterion~\eqref{eq:2fluidJS1} was formulated in  hydrodynamic approximation which mean that for stellar disc it is valid only for long-wave perturbations. In general case, it is necessary to take the collisionless Boltzmann equation for the stellar disc instead of the hydrodynamic equation. 

%%%%%%%%%%%%%%%%%%%%%%%%%%%%%%%%%%%%%%%%%%%%%%%%%%%%%%%%%%%%%%%%%%%%%
\subsection{Kinetic approximation}
%%%%%%%%%%%%%%%%%%%%%%%%%%%%%%%%%%%%%%%%%%%%%%%%%%%%%%%%%%%%%%%%%%%%%
Using collisionless Boltzmann equation for describing a stellar disc changes the dispersion relation for gravitational instability. \citet{Rafikov2001} considered this case and obtained the following expression for the criterion of two-component gravitational instability in the kinetic approximation
%%%%%%%%%%%%%%%%%%%%%%%%%%%%%%%%%%%%%%%%%%%%%%%%%%%%%%%%%%%%%%%%%%%%%
\begin{equation}
\frac{1}{Q(\bar{k})} \equiv
\frac{2}{Q_\mathrm{s}}\frac{1}{\bar{k}}
\left[1-e^{-\bar{k}^2} I_{0}(\bar{k}^2)\right] + 
\frac{2}{Q_\mathrm{g}}s\frac{\bar{k}}
{1 + \bar{k}^2 s^2} > 1 \, ,
\label{eq:rafikov}
\end{equation}
%%%%%%%%%%%%%%%%%%%%%%%%%%%%%%%%%%%%%%%%%%%%%%%%%%%%%%%%%%%%%%%%%%%%%
where $I_0$ is the modified Bessel function of the first kind.

For solving this inequality it is necessary to find all maximums of the left-hand side of the expression for given parameters $\sigma_\mathrm{s}$, $c_\mathrm{g}$, $\varkappa$, $\Sigma_\mathrm{s}$ and $\Sigma_\mathrm{g}$ or for given values $Q_\mathrm{s}$, $Q_\mathrm{g}$ and $s=\sigma_\mathrm{s}/c_\mathrm{g}$\footnote{In the kinetic approximation $\sigma_R$ should be used instead of $\sigma_\mathrm{s}$ like in one-fluid criterion \citet{Toomre1964}.}. In practice one should set to zero the derivative with respect to $\bar{k}$ of the left-hand side and solve the resulting equation numerically\footnote{The Bessel function $I_0(x)$ and its first derivative $I_1(x)$ grow as fast as ${\displaystyle \exp{x^2}}$ and hence can lead to significant errors in numerical calculations. For this reason normalized Bessel functions like \texttt{scipy.special.i0e} in \texttt{python} should be used.}. If the maximal value $Q^{-1}_\mathrm{eff}$ of the expression~\eqref{eq:rafikov} is greater than 1 then there are perturbations which make two-component disc unstable. Otherwise, the disc remains stable.

%%%%%%%%%%%%%%%%%%%%%%%%%%%%%%%%%%%%%%%%%%%%%%%%%%%%%%%%%%%%%%%%%%%%%
\subsection{Disc thickness}
%%%%%%%%%%%%%%%%%%%%%%%%%%%%%%%%%%%%%%%%%%%%%%%%%%%%%%%%%%%%%%%%%%%%%
Both criteria \eqref{eq:2fluidJS1} and \eqref{eq:rafikov} are true for infinitely thin discs. In case of a finite thickness of gaseous and stellar discs, the density of matter near $ z = 0 $ decreases and hence gravitational force decreases too. 
For this reason a thick disc becomes more stable against any perturbations. 
\citet {Jog_Solomon1984} showed how to modify the dispersion equation and the corresponding instability criterion~\eqref{eq:2fluidJS1}. The finite disc height results in the effective reduction in $\Sigma$ by a factor $\{ [1-\exp(-k h_z)]/kh_z \} $, where $2h_z$ and $k$ are respectively the total scale height and the wave number of the perturbation. Both terms at the left-hand side of the expression~~\eqref{eq:2fluidJS1} should be multiplied to this factor with $h_\mathrm{s}$ and $h_\mathrm{g}$ related to stars and gas respectively.
This recipe can be also applied to expression~\eqref{eq:rafikov}. The maximum of the corresponding expression can be found numerically.

\citet{Romeo1992,Elmegreen1995,Romeo1994,Romeo_Wiegert2011} also took into account the thickness impact. \citet{Romeo_Wiegert2011} give a simple approximation of the two-component criterion for discs of finite thickness
%%%%%%%%%%%%%%%%%%%%%%%%%%%%%%%%%%%%%%%%%%%%%%%%%%%%%%%%%%%%%%%%%%%%%
\begin{equation}
    \frac{1}{Q_\mathrm{eff, RW2}}=
    \begin{cases}
    \displaystyle
      \frac{W}{T_\mathrm{s} Q_\mathrm{s}} + \frac{1}{T_\mathrm{g} Q_\mathrm{g}},\: 
      T_\mathrm{s} Q_\mathrm{s} \ge T_\mathrm{g} Q_\mathrm{g} \, , \\[1.0em]
    \displaystyle
      \frac{1}{T_\mathrm{s} Q_\mathrm{s}} + \frac{W}{T_\mathrm{g} Q_\mathrm{g}},\: 
      T_\mathrm{g} Q_\mathrm{g} \ge T_\mathrm{s} Q_\mathrm{s} \, ,
    \end{cases}
\label{eq:RW2}
\end{equation}
%%%%%%%%%%%%%%%%%%%%%%%%%%%%%%%%%%%%%%%%%%%%%%%%%%%%%%%%%%%%%%%%%%%%%
where
%%%%%%%%%%%%%%%%%%%%%%%%%%%%%%%%%%%%%%%%%%%%%%%%%%%%%%%%%%%%%%%%%%%%%
$$
T \approx 0.8 +0.7\frac{\sigma_z}{\sigma_R} \, .
$$
%%%%%%%%%%%%%%%%%%%%%%%%%%%%%%%%%%%%%%%%%%%%%%%%%%%%%%%%%%%%%%%%%%%%%
In this expression the effect of thickness depends on the ratio of vertical to radial velocity dispersion and it can be applied for $0.5 \la {\sigma_z}/{\sigma_R} \la 1.0$.

In addition \citet{Elmegreen2011} have examined how dissipation in gas affects the stability of a disc in a two-component case and found that dissipation partially compensates the stabilizing effect of the thickness. 

We constructed our main models without taking into account the effect of the disc thickness. But then we compare them with simple fiducial models of finite thickness because the effect of the thickness can be easy quantified by using the approximation~\eqref{eq:RW2}. The effect of gas dissipation is more complex and we understand that our results for fiducial models may change, if the effect of gas dissipation is significant.

%\pagebreak{}
%\newpage
%%%%%%%%%%%%%%%%%%%%%%%%%%%%%%%%%%%%%%%%%%%%%%%%%%%%%%%%%%%%%%%%%%%%%
\section{Data}
%%%%%%%%%%%%%%%%%%%%%%%%%%%%%%%%%%%%%%%%%%%%%%%%%%%%%%%%%%%%%%%%%%%%%

In this section, we give the basic information about the required observational data, give the sources of data and describe the sample of galaxies.

%%%%%%%%%%%%%%%%%%%%%%%%%%%%%%%%%%%%%%%%%%%%%%%%%%%%%%%%%%%%%%%%%%%%%
\subsection{Observational data for two-component criterion}
%%%%%%%%%%%%%%%%%%%%%%%%%%%%%%%%%%%%%%%%%%%%%%%%%%%%%%%%%%%%%%%%%%%%%

Applying the two-component criterion to real galaxies one needs a large number of different observational data. Some of these data, i.e. long slit absorption lines spectra, are quite rare and usually available for early-type galaxies which contains a small amount of gas. The observed profiles should be long enough to avoid bulge influence, noiseless and should lie along the same axis. To find suitable objects is a very difficult task.

To conduct two-component gravitational instability analysis next data is needed:
%%%%%%%%%%%%%%%%%%%%%%%%%%%%%%%%%%%%%%%%%%%%%%%%%%%%%%%%%%%%%%%%%%%%%
\begin{itemize}
\item ``cold'' gas rotation curve $v_\mathrm{c}(R)$,
\item stellar velocity dispersion line-of-sight profile along major axis $\sigma_\mathrm{los,maj}$,
\item surface density profiles of atomic $\Sigma_{\mathrm{H\, {\sc i}}}(R)$ and molecular $\Sigma_\mathrm{H_2}(R)$ gas and information about gas sound speed,
\item surface galaxy photometry with performed decomposition into a disc and a bulge,
\item information about star-forming regions (H$\alpha$ emission, blue knots, UV and IR data).
\end{itemize}
%%%%%%%%%%%%%%%%%%%%%%%%%%%%%%%%%%%%%%%%%%%%%%%%%%%%%%%%%%%%%%%%%%%%%

It is also necessary to know the distance to the galaxy and its inclination angle $i$. All data profiles must lie along the same axis. In this work we use the major axis of a galaxy. If there is an additional information about the average stellar azimuthal velocity $\bar{v}_\varphi$ or about line-of-sight stellar velocity dispersion along the minor axis $\sigma_\mathrm{los, min}$ then it allows to get additional limits on radial velocity dispersion $\sigma_R$ values.

%%%%%%%%%%%%%%%%%%%%%%%%%%%%%%%%%%%%%%%%%%%%%%%%%%%%%%%%%%%%%%%%%%%%%
\subsection{Galaxies and data sources}
%%%%%%%%%%%%%%%%%%%%%%%%%%%%%%%%%%%%%%%%%%%%%%%%%%%%%%%%%%%%%%%%%%%%%

We collected all or almost all necessary data and conducted the two-component instability analysis using dataset for 7 spiral galaxies with main parameters given in Table~\ref{table:main_parameters}. These galaxies are early-type galaxies with intermediate inclination. Composed sample is not homogeneous by data sources, i.e. we took data from different sources. All galaxies are bright enough with maximum rotation velocity up to 380~km\,s$^{-1}$ for NGC~1167. This is an example of a galaxy with a very massive disc that can significantly  contributes to the gravitational instability.

Below we describe the main used data sources. Individual peculiarities and individual description for every galaxy are presented in Appendix A which also contains images of galaxies and data profiles in Figs.~\ref{fig:data_338}--\ref{fig:data_5533}.

Gas kinematics data $v_\mathrm{c}(R)$ for all galaxies except NGC~4258 is taken from the Westerbork H\,{\sc i} survey of spiral and irregular galaxies (WHISP, \citealp{Noordermeer_etal2005,Noordermeer_vanderHulst2007b}). We used the data for NGC~1167 from the same survey but from another source \citep{Struve_etal2010}. For NGC~4258 the central part of the gas rotation curve was drawn using $\mathrm{CO}$ data \citep{Sawada_Satoh_etal2007} while for distant regions H\,{\sc i} observations were used \citep{Eymeren_etal2011}. To reconcile data for NGC~4725 and NGC~5533 we took rotation curves from \citet{Courteau1997} (H\,{\sc i}), \citet{Eymeren_etal2011}, \citet{Zasov_etal2012} (H$\beta$, [O\,{\sc iii}] lines), \cite {Epinat_etal2008} (H$\alpha$ line).

The line-of-sight stellar velocity dispersion profile along the major axis $\sigma_\mathrm{los, maj}$ can be obtained from observations using Integral Field Unit (IFU) or a long-slit spectrograph. Despite a large number of IFU surveys appeared over the last years such as MaNGA, SAMI, CALIFA, DiskMass, $\mathrm{ATLAS^{3D}}$, almost all of them cover only central parts of a galaxy. Long $\sigma_\mathrm{los, maj}$ profiles are rare. Required data for galaxies NGC~2985, NGC~3898 and NGC~5533 were found in \citet{Noordermeer_etal2008}. These galaxies were observed using the PPAK IFU \citep{Verheijen2004}. The obtained profiles extend up to $100\arcsec$ what make them longest in the whole dataset. We also compared the profile for NGC~2985 with data obtained using ISIS spectrograph on the William Herschel Telescope \citep{Gerssen_etal2000}. Velocity dispersion profiles for  NGC~1167 \citep{Zasov_etal2008} and NGC~338 \citep{Zasov_etal2012} were observed at SCORPIO \citep{Afanasiev_Moiseev2005} focal reducer in the long-slit mode mounted at the primary focus of the 6-m telescope of the Special Astrophysical Observatory. These profiles were measured up for a distance about $50\arcsec$ from the centre. The data for NGC~1167 were also compared with points which can be extracted from CALIFA survey maps \citep{CALIFA_2017}. The profiles for NGC~4258 and NGC~4725 are available in  \citet{Heraudeau_etal1998,Heraudeau_etal1999} and were obtained using 1.93-m telescope of the Observatoire de Haute-Provence with the CARELEC long-slit spectrograph \citep{Lemaitre1990} and have last observable point at the distance about $50-60\arcsec$. The line-of-sight velocity dispersion profiles along the minor axis $\sigma_\mathrm{los,min}$ are also available for all galaxies except last two.

Surface density profiles of atomic hydrogen $\Sigma_{\mathrm{H\,\sc i}}(R)$ for all galaxies were also taken from the WHISP observations. For five galaxies these profiles were found in \citet{Noordermeer_etal2005} while the data for NGC~4258 and NGC~4725 were taken from \citet{Yim_etal2016}. For two mentioned galaxies \citet{Yim_etal2016} provide also surface density profiles of molecular gas $\Sigma_\mathrm{H_2}(R)$ obtained from BIMA \citep{BIMA2} and IRAM HERACLES \citep{IRAM} surveys. All analyzed galaxies have a large total amount of atomic hydrogen from $4\times10^9$ to $3\times10^{10}\,M_{\sun}$, but it may be distributed across large area. For example NGC~1167 shows a regular rotating 160 kpc disc with a total H{\sc i} mass equals  $1.7\times10^{10}\,M_{\sun}$, but atomic hydrogen surface density is very low (<2\,$M_{\sun}$\,pc$^{-2}$). The largest central values of $\Sigma_{\mathrm{H\,\sc i}}$ are about 10\,$M_{\sun}$\,pc$^{-2}$ in the NGC~338 and the lowest are in NGC~1167 and NGC~3898. Besides NGC~4258 and NGC~4725 only NGC~2985 has available observational data for the molecular gas surface density profile, taken from \citet{Young_etal1995}. For galaxies NGC~338 \citep{Lavezzi_etal1998}, NGC~1167 \citep{OSullivan_etal2015} and NGC~3898 \citep{Boselli_etal2014} we found information about the intensity of $^{12}\mathrm{CO}(J=1\rightarrow 0)$ line which allows to calculate the total mass of molecular hydrogen and to build a model-dependent surface density profile. Details about these model profiles and a NGC~5533 separate case are described in the next section.

The information about photometry and bulge-disc decomposition was taken from  \citet{Noordermeer_vanderHulst2007} in the $B,\,R,\,I$ bands; \citet{Mollenhoff_Heidt2001} in the $J,\,H,\,K$ bands; \cite {Gutierrez_etal2011} in the $R$-band; \citet{Mendez_etal2008} in the $J$-band; \citet{S4G,Fisher_Drory2010} in the $3.6\,\micron$ band. For the galaxy NGC~4258 we use the data in the $V,\,I,\,J$ bands \citep{Yoshino_Ichikawa2008}.

The presence and the size of the large-scale star formation regions were visually selected from blue knots distribution in images from Sloan Digital Sky Survey (SDSS) as well as from maps in H$\alpha$ emission  \citep{Hameed_Devereux2005,Epinat_etal2008,CALIFA_2016}, UV emission according to GALEX and IR emission according to SPITZER. The SFR in these galaxies varies within the range $1-5\, M_{\sun}/$yr \citep{DiTeodoro_Fraternali2014,Theios_etal2016}. The only exception is NGC~1167 where the birth rate of new stars is rather small and equals to 0.3 $M_{\sun}/$yr \citep{CALIFA_2016}.

In this paper we use the Hubble constant  $H_\mathrm{o}=75\:\mathrm{km\,s^{-1}\,Mpc^{-1}}$. All data from the literature were corrected to this accepted value.

%%%%%%%%%%%%%%%%%%%%%%%%%%%%%%%%%%%%%%%%%%%%%%%%%%%%%%%%%%%%%%%%%%%%%
\begin{table*}
%%%%%%%%%%%%%%%%%%%%%%%%%%%%%%%%%%%%%%%%%%%%%%%%%%%%%%%%%%%%%%%%%%%%%
\caption{The main parameters of the examined galaxies.}
\label{table:main_parameters}
\begin{tabular}{cccccccc}
\hline
Galaxy & Type & Inclination $i$, degr & Distance, Mpc & Scale, $\mathrm{kpc}/\arcsec$ & $M_\mathrm{B},{\rm{mag}}$ & $M_{\mathrm{H\, {\sc i}}}, \, 10^9\times M_{\sun}$ & $M_\mathrm{H_2}, \, 10^9\times M_{\sun}$\\
(1) & (2) & (3) & (4) & (5) & (6) & (7) & (8)\\
\hline
NGC~338 & Sab & $64\pm4^{\rm a,b,c}$ & 65.1 & 0.316 & $-21.5$ & 14.42 & 5.12 \\ 
NGC~1167 & SA0 & $38\pm 2^{\rm b,d,e,f}$ & 67.4 & 0.327 & $-21.7$ & 17.09 & 1.85\\ 
NGC~2985 & (R)SA(r)ab & $36\pm 2^{\rm g,h,b,i,f,j,k}$ & 21.1 & 0.102 & $-20.9$ & 13.92 & $1.91^*$\\ 
NGC~3898 & SA(s)ab & $61\pm 8^{\rm g,b,i,j,l}$ & 18.9 & 0.092 & $-20.7$ & 3.96 & 0.20\\
NGC~4258& SABb & $65\pm 5^{\rm m,n,o,k,p}$ & $7.9^*$ & 0.038 & $-20.9$ & $8.20^*$ & $1.07^*$\\
NGC~4725& SABa & $50\pm 6^{\rm b,m,n,k}$ & 18.2 & 0.088 & $-20.7$ & 9.78 & $2.49^*$\\
NGC~5533& SA(rs)ab & $52\pm 1^{\rm b,m,f}$ & 54.3 & 0.263 & $-21.5$ & 30.23 & $8.11^*$\\
\hline 
\end{tabular}
\\
%%%%%%%%%%%%%%%%%%%%%%%%%%%%%%%%%%%%%%%%%%%%%%%%%%%%%%%%%%%%%%%%%%%%%
Morphological type ${\bf(2)}$ was taken from NASA/IPAC Extragalactic Database (NED). Inclination angles ${\bf(3)}$ were collected from various papers: (a)~\citet{Zasov_etal2012} (b)~\citet{Noordermeer_etal2005} (c)~\citet{Courteau1997} (d)~\citet{Zasov_etal2008} (e)~\citet{CALIFA_2016} (f)~\citet{Noordermeer_vanderHulst2007}
(g)~\citet{Noordermeer_etal2008} (h)~\citet{Dumas_etal2007} (i)~\citet{Epinat_etal2008} (j)~\citet{Gutierrez_etal2011} (k)~\citet{S4G} (l)~\citet{Mollenhoff_Heidt2001} (m)~\citet{Eymeren_etal2011} (n)~\citet{Yim_etal2016} (o)~\citet{Sawada_Satoh_etal2007} (p)~\citet{Yoshino_Ichikawa2008}. 
Distances to galaxies ${\bf(4)}$ and scales ${\bf(5)}$ were found in \citet{Noordermeer_etal2005,Yim_etal2016} and corrected assuming $H_\mathrm{o}$ = 75~$\mathrm{km\,s^{-1}\,Mpc^{-1}}$. The absolute magnitude in the $B$-band ${\bf(6)}$ was taken from the LEDA database (Lyon-Meudon Extragalactic Database). Hydrogen total mass ${\bf(7)}$ data were found in \citet{Noordermeer_etal2005}. Total $M_\mathrm{H_2}$ masses ${\bf(8)}$ were found in related sources (see text), corrected for uniformity according to \citet{Yim_etal2016} to $X_{\mathrm{CO}} = 1.9\times 10^{20}\, \mathrm{cm^{-2}\, K^{-1}\, km^{-1}}$ \citep{Strong_Mattox1996} and adjusted to distances (4). In cases marked by asterisk $^*$  gas masses were obtained during direct numerical integration of surface densities at available radii and should be considered as mass within a few exponential scale lengths. 
\end{table*}
%%%%%%%%%%%%%%%%%%%%%%%%%%%%%%%%%%%%%%%%%%%%%%%%%%%%%%%%%%%%%%%%%%%%%

% \pagebreak{}
%\newpage
%%%%%%%%%%%%%%%%%%%%%%%%%%%%%%%%%%%%%%%%%%%%%%%%%%%%%%%%%%%%%%%%%%%%%
\section{Method}
%%%%%%%%%%%%%%%%%%%%%%%%%%%%%%%%%%%%%%%%%%%%%%%%%%%%%%%%%%%%%%%%%%%%%

In this section we describe details of performed analysis, equations solving and observational data processing.  

%%%%%%%%%%%%%%%%%%%%%%%%%%%%%%%%%%%%%%%%%%%%%%%%%%%%%%%%%%%%%%%%%%%%%
\subsection{Kinematics}
%%%%%%%%%%%%%%%%%%%%%%%%%%%%%%%%%%%%%%%%%%%%%%%%%%%%%%%%%%%%%%%%%%%%%

The ``cold'' gas rotation curve $v_\mathrm{c}(R)$ was corrected for systematic velocity, bend across the galaxy centre and corrected for inclination by factor $\sin i$. The resulting profile fitted by cubic smoothing B-spline curve. The epicyclic frequency $\varkappa$ for an infinitely thin disc can be find as follows
$$\varkappa = \sqrt{2}\frac{v_\mathrm{c}}{R}\sqrt{1+\frac{R}{v_\mathrm{c}}\frac{dv_\mathrm{c}}{dR}}.$$
The derivative in the formula above was calculated numerically and verified by $\varkappa$ recalculation for fitting of the rotation curve by a small degree polynomial. In case of NGC~338 where the rotation curve is short we calculated the epicyclic frequency as $\varkappa = \displaystyle \sqrt{2}\frac{v_\mathrm{c}}{R}$ for distances greater than 60\arcsec assuming $v_\mathrm{c}(R)$ to be flat there.

It is difficult to recover the stellar velocity ellipsoid (SVE) components from observational data since this task belongs to the class of incorrect problems where more than one solution can be found \citep{Gerssen_etal1997,Gerssen_etal2000,Shapiro_etal2003,Marchuk_Sotnikova2017}. two-component instability analysis requires only one of the components namely the stellar velocity dispersion in the radial direction $\sigma_R$. For all galaxies we found profiles of the stellar line-of-sight velocity dispersion along the major axis $\sigma_\mathrm{los,maj}$ which are connected to the required component by the following equation
%%%%%%%%%%%%%%%%%%%%%%%%%%%%%%%%%%%%%%%%%%%%%%%%%%%%%%%%%%%%%%%%%%%%%
\begin{equation}
\displaystyle 
\sigma_\mathrm{los,maj}^2 = 
\sigma_R^2\left(\frac{\sigma_\varphi^2}{\sigma_R^2}\,
\sin^2 i+\frac{\sigma_z^2}{{\sigma_R^2}}\,\cos^2 i\right)\, ,
\label{eq:sve} 
\end{equation}
%%%%%%%%%%%%%%%%%%%%%%%%%%%%%%%%%%%%%%%%%%%%%%%%%%%%%%%%%%%%%%%%%%%%%
where $\sigma_\varphi$ and $\sigma_z$ are azimuthal and vertical components of the SVE respectively. It is impossible to extract the exact value of $\sigma_R$ from Eq.~\eqref{eq:sve}, but with some additional conditions we can obtain constraints on $\sigma_R$.

First additional condition is valid for an equilibrium disc in the epicyclic approximation \citep{Binney_Tremaine2008}:
%%%%%%%%%%%%%%%%%%%%%%%%%%%%%%%%%%%%%%%%%%%%%%%%%%%%%%%%%%%%%%%%%%%%%
\begin{equation}\label{eq:phi_to_R} 
\frac{\sigma_\varphi^2}{\sigma_R^2} = 
0.5\left(1+\frac{\partial\ln v_\mathrm{c}}{\partial\ln R}\right)\,.
\end{equation}
%%%%%%%%%%%%%%%%%%%%%%%%%%%%%%%%%%%%%%%%%%%%%%%%%%%%%%%%%%%%%%%%%%%%%
We use this equation to estimate the first term in the Eq.~\eqref{eq:sve}. It is possible to calculate the derivative $\displaystyle \frac{\partial\ln v_\mathrm{c}}{\partial\ln R}$ directly, but in practice the result depends on approximations and unstable because of the multiplier $R$ in the derivative. In the epicyclic approximation the term $(\sigma_\varphi/\sigma_R)^2$ can be written as $(\varkappa/2\Omega)^2$, where $\varkappa(R)$ and $\Omega(R)=v_\mathrm{c}/R$ are easily computed. Unfortunately the result is also very waved due to the same factor $R$, although it lies almost entirely between two bounds that we estimated as follows. In central regions of a galaxy the disc rotates solidly thus the expression in parentheses in Eq.~\eqref{eq:phi_to_R} is equal to 2 while at the periphery the rotation curve remains flat and its derivative is close to zero. Hence we can use the inequality $0.5 \le (\sigma_\varphi/\sigma_R)^2 \le 1$.

The second term in Eq.~\eqref{eq:sve} can be limited as $0.3 \le \sigma_z/\sigma_R \le 0.7$, where the lower limit supports the disc against the bending instability \citep{Rodionov_Sotnikova2013} and the upper one comes from the analysis of observational data.  
\citet{Gerssen_Shapiro2012} give the upper limit of about 1 and demonstrate the trend of $\sigma_z/\sigma_R$ with a galaxy type with $\sigma_z/\sigma_R=1$ for the early type galaxies. 
Note that the only galaxy in figure~4 of \citet{Gerssen_Shapiro2012} has $\sigma_z/\sigma_R <1.02 \pm 0.11$. It is the case of NGC 2775. 
As we have shown in \citet{Marchuk_Sotnikova2017} this result is incorrect. For NGC 2775 it is impossible to recover SVE correctly because high inclination of the galaxy makes the contribution of the vertical velocity dispersion component to the line-of-sight velocity dispersion data comparable with observational
uncertainties that leads to the degeneracy of the solution. As a result the authors obtained a formal optimal value of $\sigma_z/\sigma_R$ that accidentally fall at the upper boundary of the range of $\sigma_z/\sigma_R$. \citet{Westfall_etal2014}, 
averaging over their subsample of late-type galaxies, gave a meridional shape of $\sigma_z/\sigma_R=0.51$. \citet{Pinna_etal2016} constrained the SVE for a sample of galaxies from CALIFA survey \citep{CALIFA_2017} and found no trend of $\sigma_z/\sigma_R$ with the galaxy type with typical values of $\sigma_z/\sigma_R=0.7$.  Moreover \citet{Marchuk_Sotnikova2017} obtained for a lenticular galaxy NGC~1167 $\sigma_z/\sigma_R=0.3$ in the outer regions of the stellar disc and $\sigma_z/\sigma_R=0.7$ in the inner parts. For all these reasons 
we consider the range $\displaystyle 0.3 \le \sigma_z/\sigma_R \le 0.7$.

Both unequalities substituting in the Eq.\eqref{eq:sve} give:
%%%%%%%%%%%%%%%%%%%%%%%%%%%%%%%%%%%%%%%%%%%%%%%%%%%%%%%%%%%%%%%%%%%%%
\begin{equation}
\frac{\sigma_\mathrm{los,maj}}{\sqrt{\sin^2 i + 0.49\cos^2 i}}< \sigma_R < \frac{\sigma_\mathrm{los,maj}}{\sqrt{0.5\sin^2 i + 0.09\cos^2 i}}\,.\label{eq:sigR}
\end{equation}
%%%%%%%%%%%%%%%%%%%%%%%%%%%%%%%%%%%%%%%%%%%%%%%%%%%%%%%%%%%%%%%%%%%%%

For NGC~1167 both limits are consistent with values founded in \citet{Marchuk_Sotnikova2017} where we used additional  observational data along the minor axis $\sigma_\mathrm{los, min}$. In the following analysis both derived limits on $\sigma_R$ were used to find the upper and the lower values of $Q_\mathrm{eff}$ which bound the area of actual $Q_\mathrm{eff}$ values. 

To compute boundaries for $\sigma_R$ we bend across the centre the observed profile $\sigma_\mathrm{los,maj}(R)$ and fit it by smoothed cubic splines multiplied by the corresponding constant from \eqref{eq:sigR}. In order to exclude the bulge influence we did not consider all points closer to the centre than the one effective radius of the bulge $r_\mathrm{e, b}$. If several decomposition parameters are available the largest $r_\mathrm{e, b}$ is used. Radial velocity dispersion profiles were stretched to a distant regions without observable points at the same level as the last observed point. Since the real value of $\sigma_R$ at the periphery tends to decrease with increasing distance, this approximation results in more stable two-component disc. At the same time, we did not consider areas too far from the last observed point of $\sigma_\mathrm{los,maj}$.

Usually one takes the velocity gas dispersion $\sigma_\mathrm{g}$ instead of the sound speed $c_\mathrm{g}$ \citep{Schaye2004}, because it can be obtained from observations. Such radial profiles of $\sigma_\mathrm{g}(R)$ for atomic and molecular hydrogen can be found in \citet{Boomsma_etal2008,Tamburro_etal2009,Ianjamasimanana_etal2012,Zheng_etal2013,Mogotsi_etal2016,Romeo_Mogotsi2017}. Unfortunately, galaxies studied in this paper are not among those listed in referred works. For this reason we use a constant value of $\sigma_\mathrm{g}$ value in our analysis. \citet{Kennicutt1989} adopted $\sigma_\mathrm{g} = 6 $~km\,s$^{-1}$ while \citet{Leroy_etal2008} used $\sigma_\mathrm{g} = 11 $~km\,s$^{-1}$. These are in good agreement  with gas observations in the Galaxy and profiles from \citet{Mogotsi_etal2016}. Note that average dispersion of atomic hydrogen is greater than average dispersion of molecular gas. In this paper we assume gas velocity dispersion $\sigma_\mathrm{g} = 6$~km\,s$^{-1}$ for both gaseous components. We have tested this assumption and measured how a change in $\sigma_\mathrm{g}$ affects the final result below.

%%%%%%%%%%%%%%%%%%%%%%%%%%%%%%%%%%%%%%%%%%%%%%%%%%%%%%%%%%%%%%%%%%%%%
\subsection{Gas distribution}
%%%%%%%%%%%%%%%%%%%%%%%%%%%%%%%%%%%%%%%%%%%%%%%%%%%%%%%%%%%%%%%%%%%%%

For all galaxies the extended profiles of $\Sigma_\mathrm{H\,\sc i}$ obtained from observations at the WSRT telescope were found in \citet{Yim_etal2016,Noordermeer_etal2005}. The surface density of $\Sigma_\mathrm{H_2}$ is usually calculated from observations of $\mathrm{CO}$ molecule lines, but the exact value of the conversion factor $X_\mathrm{CO}$ remains debatable. If $\mathrm{CO}$ intensity profile is available, we follow the procedure described in \citet{Yim_etal2016} to obtain $\Sigma_\mathrm{H_2}$. For NGC~4258, NGC~4725 \citep{Yim_etal2016} and NGC~2985 \citep{Young_etal1995} the $\mathrm{CO}(J=1\rightarrow 0)$ line was used to calculate $\Sigma_\mathrm{H_2}$:
$$\Sigma_\mathrm{H_2} [M_{\sun}\,\mathrm{pc}^{-2}] = 
3.2\times I_\mathrm{CO}[\mathrm{K\,km\,s^{-1}}]\,.$$

However, for three of four remaining galaxies only the total mass of molecular hydrogen $M_\mathrm{H_2}$ was found. \citet{BIMA1} make a conclusion based on the BIMA survey data that  $\Sigma_\mathrm {H_2}(R)$ profile decreases exponentially with a scale equal to the scale length of the disc $h$ on the average. Therefore for these remaining galaxies we found central surface densities from the formula of the total mass of the molecular disc $\displaystyle \Sigma_\mathrm{H_2}(0) = \frac{M_\mathrm{H_2}}{2\pi \, h^2}$ and then extended them exponentially with the appropriate $h$ scale length. If more than one photometry is used for an examined galaxy this approach gives several possible profiles of $\Sigma_\mathrm{H_2}(R)$ and several estimates of $Q$, respectively both for one-fluid and two-component cases. We approximate the $\Sigma_\mathrm{H_2}(R)$ and $\Sigma_\mathrm{H\, \sc i}(R)$ profiles using linear interpolation between points. The total gas surface density was corrected for the presence of helium and other heavy elements using 
$\Sigma_\mathrm{g} = 
1.36\,(\Sigma_\mathrm{H\,\sc i} + \Sigma_\mathrm{H_2})$ formula  (see, for example, \citealp{Leroy_etal2008}).

For NGC~5533 we did not find any data about molecular hydrogen and used another technique for constructing $\Sigma_\mathrm{g}$. \citet{Yim_etal2016} showed (see their figure 3) that central values of $\Sigma_\mathrm{g}$ on average can be obtained as an exponential continuation of the $\Sigma_{\mathrm{H\,{\sc i}}}$ profile toward the galactic centre with a scale of $r_{25}/1.92$. We fit $\Sigma_{\mathrm{H\, {\sc i}}}$ profile for NGC~5533 using this scaling relation within $1.0-2.0 \, r_{25}$, where this relation is most accurate according to the original figure and where molecular hydrogen contribution in the total profile is negligible ($r_{25}=78 \arcsec$ in $B$-band). We continued this fit to the centre and found the central surface density $\Sigma_\mathrm{g}$ to be around 27~$M_{\sun}$\,pc$^{-2}$. Values of $\Sigma_\mathrm{H_2}$ profile were obtained as the difference between this exponential approximation $\Sigma_\mathrm{g}$ and the observed $\Sigma_{\mathrm{H\, {\sc i}}}$. As before, the total gas surface density was corrected for the presence of helium. The constructed $\Sigma_\mathrm{g}$ profile is strongly model dependent but the usage of another scaling relation \citet{Bigiel_Blitz2012} or $r_{25}$ for another photometry does not affect the results much. The method predicts a large total amount of molecular gas in the galaxy and it is unnatural that NGC~5533 was not included in any $\mathrm{CO}$ survey. For this reason we also analyzed the additional model without taking into account the molecular gas, where $\Sigma_\mathrm{g} = 1.36 \times \Sigma_{\mathrm{H\, {\sc i}}}$.

%%%%%%%%%%%%%%%%%%%%%%%%%%%%%%%%%%%%%%%%%%%%%%%%%%%%%%%%%%%%%%%%%%%%%
\subsection{Disc mass model}
%%%%%%%%%%%%%%%%%%%%%%%%%%%%%%%%%%%%%%%%%%%%%%%%%%%%%%%%%%%%%%%%%%%%%

Surface photometry of the galaxy along with the decomposition into a bulge and disc were used to obtain a stellar disc mass model and a $\Sigma_\mathrm{s}(R)$ profile. Initially for each galaxy we found a large number of such decompositions in optical and infrared bands (minimum 3 bands for NGC 338, maximum 9 bands for NGC~2985 and NGC~5533). Values of $M/L$ were derived from statistical calibration relations in \citet{Bell_etal2003,McGaugh_Schombert2014,Querejeta_etal2015}: 
$$\log_{10}{M/L} = a_\lambda + (b_\lambda \times \mathrm{Colour})\,,$$
where $a_\lambda$ and $b_\lambda$ are numerical coefficients for the $\lambda$-band, $\mathrm{Colour}$ is a disc colour. As $\mathrm{Colour}$ we took the $B-R$ colour from the data in  \citet{Noordermeer_vanderHulst2007} or all possible colours for SDSS bands\footnote{In this case absolute stellar magnitudes of the disc were corrected for the  FOREGROUND GALACTIC EXTINCTION taken from NED.} or the difference $3.6\, \micron - 4.5 \,\micron$ in case of photometry from $S^4G$ survey. All magnitudes were corrected for absorption, where possible. Magnitudes in $S^4G$ are in AB-values and were converted to Vega magnitudes and then to luminosities. In case of two-disc model the total surface density $\Sigma_\mathrm{s} $ was equal to the sum of both individual stellar disc profiles.

%%%%%%%%%%%%%%%%%%%%%%%%%%%%%%%%%%%%%%%%%%%%%%%%%%%%%%%%%%%%%%%%%%%%%
\begin{figure}
\includegraphics[width=\columnwidth]{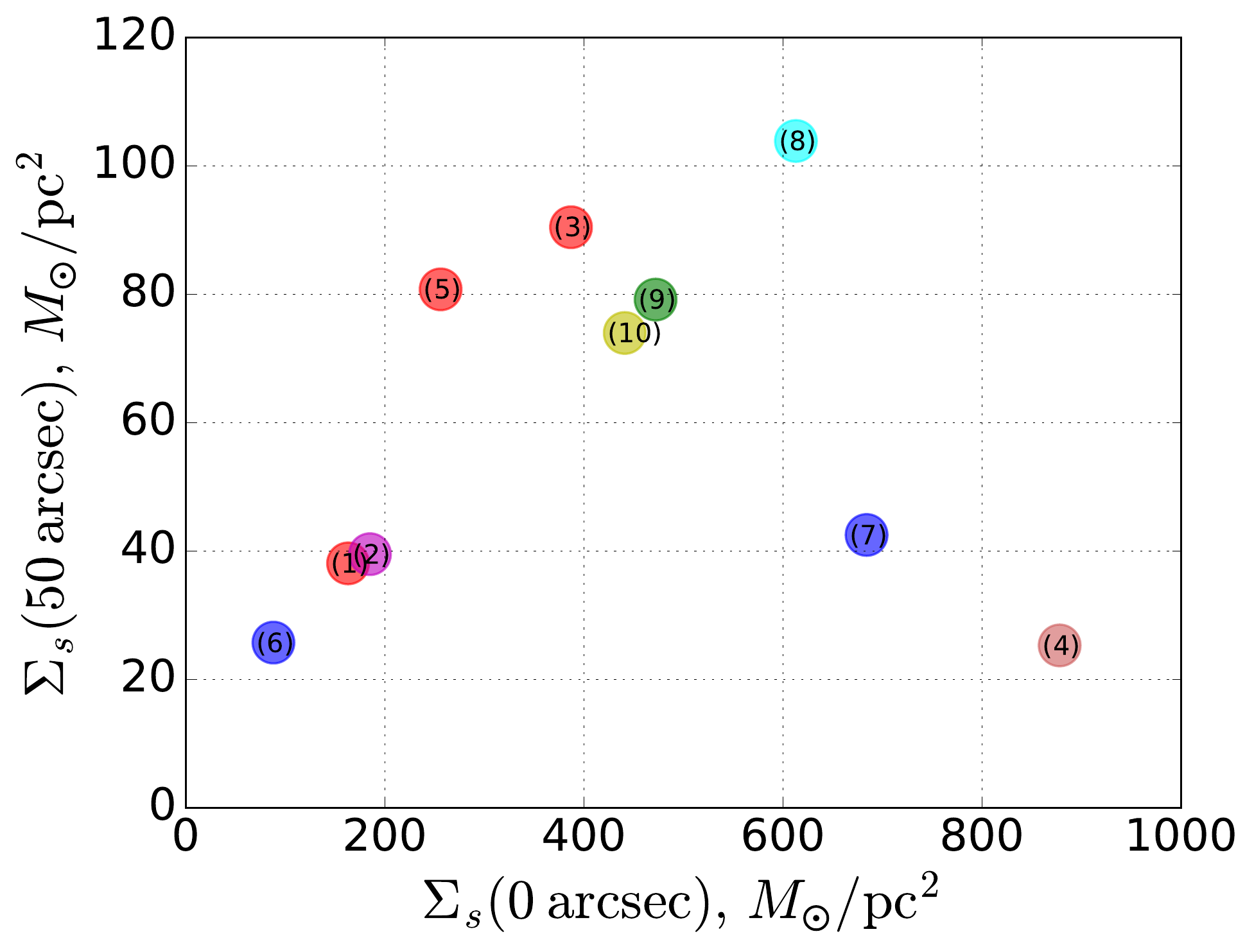}
\caption{Mass models of the NGC~5533 stellar disc, obtained from the calibration relations \citet{Bell_etal2003, McGaugh_Schombert2014, Querejeta_etal2015}. The dependence of the surface density $\Sigma_\mathrm{s}$ in the centre of the galaxy and at the distance of $50\arcsec$ is shown. Marker's colour represent photometric band. Photometries used are \citet{Noordermeer_vanderHulst2007}
(1) $R$; (2) $B$; (3) maximal disc $R$; (4) \citet{Mendez_etal2008} $J$; (5) \citet{Broeils_Knapen1991} $R$; (6) \citet{Baggett_etal1998} $V$; (7) \citet{Silchenko_etal1998} $V$; \citet{Mendez_etal2017} (8) $g$ ($r-i$); (9) $r$ ($g-i$); (10) $i$ ($g-r$). For all cases except (8)-(10) we use corrected for extinction $B-R$ colour from  \citet{Noordermeer_vanderHulst2007}. For three last photometries used colours were calculated from decomposition data in \citet{Mendez_etal2017} and showed in brackets.}
\label{fig:5533_phot_variance} 
\end{figure}
%%%%%%%%%%%%%%%%%%%%%%%%%%%%%%%%%%%%%%%%%%%%%%%%%%%%%%%%%%%%%%%%%%%%%

Unfortunately mass models for individual galaxies using different photometry and decomposition were inconsistent and the central value of $\Sigma_\mathrm{s}$ may vary by a factor of five (Fig.~\ref{fig:5533_phot_variance}). This is due to the statistical nature of the calibration relations, which allow individual deviations from the mean value, and because of insufficient calibration of different bands between themselves \citep{McGaugh_Schombert2014}. Therefore we decided to use a maximal disc model for decompositions in bands that are closest to the infrared region of the spectrum as they are less contaminated by dust. We define a maximal disc model as the model with the largest value of $M/L$ for a given photometry which contribution to the rotation curve  $v_\mathrm{c}$ does not exceed $0.85 \, v_\mathrm{c,max}$ \citep{Sackett1997,Courteau_etal1999}. We also used the sub-maximal disc model with corresponding restriction of rotation curve equal $0.6 \, v_\mathrm{c,max}$ \citep{Courteau_etal1999} for NGC~338 and NGC~1167 using decomposition in the $B$-band as $M/L$ in this band turned out to be too large for a maximal model. 

All photometries, their decomposition parameters and $M/L$ values chosen for further analysis are shown in Table~\ref{table:photo_parameters}. We also present  corresponding rotation curves for picked mass models for individual galaxies in Figs.~\ref{fig:data_338}--\ref{fig:data_5533}. Note that maximal disc assumption leads to overestimation of $\Sigma_ \mathrm{s}$ values and results in maximal destabilizing perturbation that stars can exert on gas in a two-component model. The choice of maximal disc model can be justified by fact that all galaxies are bright (see Table~\ref{table:main_parameters}) and that they have very high rotational velocities, which is evidence in favor of massive stellar discs as it can be seen from the final estimates of $M/L$ ratios in Table~\ref{table:photo_parameters}.

%%%%%%%%%%%%%%%%%%%%%%%%%%%%%%%%%%%%%%%%%%%%%%%%%%%%%%%%%%%%%%%%%%%%%
\subsection{Star formation indicators}
%%%%%%%%%%%%%%%%%%%%%%%%%%%%%%%%%%%%%%%%%%%%%%%%%%%%%%%%%%%%%%%%%%%%%
We use several indicators of star formation, mostly the presence of blue knots in images from SDSS survey and $H{\alpha}$ emission along with UV and IR emission in some special cases. The exact radial profile of SFR was not analyzed, instead we measured distances from the centre to the outer regions beyond which there is no active star formation or its rate significantly reduced. For all galaxies except NGC~4725 it is possible to find such boundary distance and the related area which contains significant star formation at any radius and azimuth. NGC~4725 contains radial star-forming structures visible in IR (see Appendix A) which can not be azimuthally averaged. For NGC~1167 where a weak star formation activity is concentrated mostly in the spirals-like rings, the examined SFR region was also constrained from inside by a minimal distance. For all measurements we use distances to galaxies $D$ given in Table~\ref{table:main_parameters}. It should be noted that $D$ measured in other works may differ by factor of two from those used here, like for NGC~4725 \citep{Hameed_Devereux2005}. 
For two galaxies with massive outer spirals NGC~4258 and NGC~4725 we also determined the extension of these spirals but their remoteness from the centre allows to perform only single-fluid instability analysis (see Fig.~\ref{fig:outer_spirals}). Thus our star formation analysis is rather qualitative than quantitative but it is still sufficient for instability analysis.

%%%%%%%%%%%%%%%%%%%%%%%%%%%%%%%%%%%%%%%%%%%%%%%%%%%%%%%%%%%%%%%%%%%%%
\begin{table*}
\caption{Photometric and mass parameters of the examined galaxies.}
\label{table:photo_parameters}
\begin{tabular}{ c c c c c c l c }
\hline
Galaxy & Band & Source & $r_\mathrm{e, b}$, arcsec  & $h$, arcsec & $\mu_0^{\mathrm{d}}, \mathrm{mag/arcsec^2}$ & M/L, $M_{\sun}/L_{\sun}$ & $M_\mathrm{d}, \, 10^{10}\times M_{\sun}$\\
(1) & (2) & (3) & (4) & (5) & (6) & \qquad(7) & (8)\\
\hline
NGC~338 & $R$ & N07 & 15.0 & 18.3 & 21.92 & \qquad$9.64^{\mathrm{sub}}$ & 8.54\\ 
   ${}$ & $B$ & N07 & 15.0 & 17.7 & 22.53 & \qquad$6.48^{\mathrm{sub}}$ & 8.16\\
\hline
NGC~1167 & $R$ & N07 & 6.7 & 24.2 & 20.12 & \qquad4.95 & 43.47\\
${}$ & --- \raisebox{-0.5ex}{''} --- & --- \raisebox{-0.5ex}{''} --- & --- \raisebox{-0.5ex}{''} --- & --- \raisebox{-0.5ex}{''} --- & --- \raisebox{-0.5ex}{''} --- & \qquad$2.47^{\mathrm{sub}}$ & 21.69\\
\hline
NGC~2985 & $K$ & H01 & 14.0 & 31.1 & 17.32 & \qquad1.27 & 8.32\\
    ${}$ & $3.6\, \micron$ & S4G & 6.3 & 12.8/48.9 & 18.55/20.84 & \qquad1.21 & 7.77\\
\hline
NGC~3898 & $R$ & N07 & 8.8 & 36.2 & 20.49 & \qquad7.80 & 8.64\\
    ${}$ & $R$ & G11 & ${-}$ & 19.11/59.9 & 19.03/21.53 & \qquad2.93 & 6.88\\
\hline
NGC~4258 & $I$ & Y08 & 14.9 & 74.2 & 18.26 & \qquad1.00 & 4.52\\
    ${}$ & $3.6\, \micron$ & F10 & 15.0 & 80.7 & 18.82 & \qquad0.38 & 10.66\\
\hline
NGC~4725 & $H$ & H01 & ${-}$ & 50.28 & 17.11 & \qquad0.72 & 11.48\\
    ${}$ & $3.6\, \micron$ & S4G & 10.14 & 73.2 & 20.34 & \qquad1.61 & 18.57\\
\hline
NGC~5533 & $r$ & M17 & 8.9 & 28.0 & 20.72 & \qquad4.51 & 24.95\\
    ${}$ & $R$ & N07 & 9.9 & 34.4 & 21.27 & \qquad7.32 & 29.93\\
\hline
\end{tabular}
\\

NGC number (1), band (2), source of photometric parameters: N07 --- \citet{Noordermeer_vanderHulst2007}; H01 --- \citet{Mollenhoff_Heidt2001}; S4G --- \citet{S4G}; G11 --- \citet{Gutierrez_etal2011}; Y08 --- \citet{Yoshino_Ichikawa2008}; F10 --- \citet{Fisher_Drory2010}; M17 --- \citet{Mendez_etal2017}, the effective bulge radius (4), the exponential disc scale length (5) and the central surface brightness (6); for galaxies with two stellar discs their photometic parameters showed divided by slash.  Estimated the mass-to-light $M/L$ ratio are given in (7) column, where $^{\mathrm{sub}}$ denoted sub-maximal mass model used. Total disc (discs) mass (8). Note that difference in $M/L$ estimation for NGC~3898 is due to different values of estimated central surface brightness.
\end{table*}
%%%%%%%%%%%%%%%%%%%%%%%%%%%%%%%%%%%%%%%%%%%%%%%%%%%%%%%%%%%%%%%%%%%%%

%%%%%%%%%%%%%%%%%%%%%%%%%%%%%%%%%%%%%%%%%%%%%%%%%%%%%%%%%%%%%%%%%%%%%
\subsection{Solution of equations}
%%%%%%%%%%%%%%%%%%%%%%%%%%%%%%%%%%%%%%%%%%%%%%%%%%%%%%%%%%%%%%%%%%%%%

\begin{figure*}
\includegraphics[width=2\columnwidth]{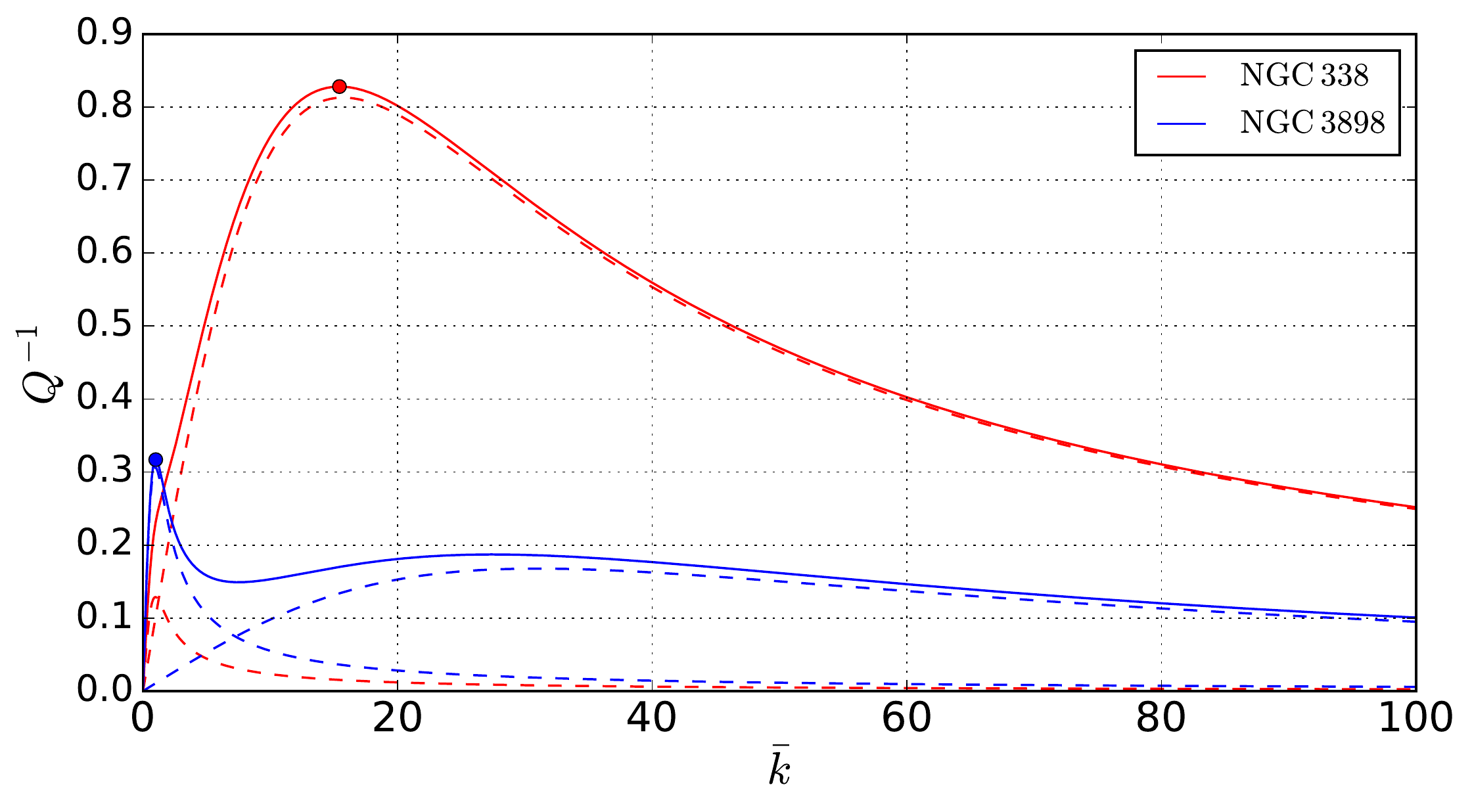}
\caption{Dependence of $Q^{- 1}$ on the dimensionless wave number $\bar{k}$. The solid line shows the dependence~\eqref{eq:rafikov} for NGC~338 (red, above) and NGC~3898 (blue, below) at distance $R=60\arcsec$ from the centre. Both models are for $R$-band photometry and upper $\sigma_R$ limit. The dashed lines show individual contribution of two terms from~\eqref{eq:rafikov} in $Q^{-1}$. The first peak corresponds to the contribution of the stellar component and the second one corresponds to the gaseous component, respectively. The dots indicate maxima which show the most unstable wave number for the two-component model. At the examined distance the most unstable component for NGC~338 is a gaseous disc  whereas for NGC~3898 it is a stellar disc.}
\label{fig:338_vs_3898_Qeff} 
\end{figure*}

We apply the instability criterion at each observational point of the $\Sigma_\mathrm{g}$ profile within a certain range of distances beyond the bulge and not so far away from the last $\sigma_\mathrm{los, maj}$ observational point. For every mass models of the disc we calculate $Q_\mathrm{g}$ and two values of $Q_\mathrm{s}$ namely $Q_\mathrm{s,max}$ and $Q_\mathrm{s,min}$, which correspond to the lower and the upper limits of the radial velocity dispersion $\sigma_R$. After that we search the maximum of the expression~\eqref{eq:rafikov} by applying either simple lookup through the grid $\bar{k}$ or solving numerically the equation in which derivative by $\bar{k}$ is equal to zero. In the last case we apply the dichotomy method on the grid using the idea that if there are different signs of the function at the ends of the grid then the interval must contain a root. The maximum of the two-component instability equation in the hydrodynamic approximation \eqref{eq:2fluidJS1} was also found for verification. This method is simpler since the corresponding equation for the derivative is cubic \citep{Elmegreen1995} and can be solved analytically. For all galaxies and observational points the difference between $Q_\mathrm{eff}$ for hydrodynamic and kinetic approximations is rather small. Hereafter we use $Q_\mathrm{eff}$ for kinetic approximation as the more precise one. Note that we solve each equation twice for the lower and upper bounds of $\sigma_R$. Fig.~\ref{fig:338_vs_3898_Qeff} shows an example of $Q^{-1}(\bar{k})$ at the same distance in galaxies NGC~338 and NGC~3898, as well as the position of the maximum and separate contributions of terms in the Eq.~\eqref{eq:rafikov}. It is seen that the maximum corresponds to the most unstable wavelength. The figure also shows which of the components, gas or stellar (second and first peak, respectively) is least stable. Finally, the Toomre parameter for the gaseous disc $Q_\mathrm{g}$, upper and lower profiles $Q_\mathrm{eff, max}$ and $Q_\mathrm{eff, min}$ for the two-component model were compared with unity in case of axisymmetric perturbations or with the threshold value $\alpha^{-1}$ in case of nonaxisymmetric perturbations. This allowed us to identify the regions of instability.

%%%%%%%%%%%%%%%%%%%%%%%%%%%%%%%%%%%%%%%%%%%%%%%%%%%%%%%%%%%%%%%%%%%%%
%%%%%%%%%%%%%%%%%%%%%%%%%%%%%%%%%%%%%%%%%%%%%%%%%%%%%%%%%%%%%%%%%%%%%

%\pagebreak{}
%\newpage
%%%%%%%%%%%%%%%%%%%%%%%%%%%%%%%%%%%%%%%%%%%%%%%%%%%%%%%%%%%%%%%%%%%%%
\section{Results and discussion}
%%%%%%%%%%%%%%%%%%%%%%%%%%%%%%%%%%%%%%%%%%%%%%%%%%%%%%%%%%%%%%%%%%%%%

%%%%%%%%%%%%%%%%%%%%%%%%%%%%%%%%%%%%%%%%%%%%%%%%%%%%%%%%%%%%%%%%%%%%%
\subsection{Results}
%%%%%%%%%%%%%%%%%%%%%%%%%%%%%%%%%%%%%%%%%%%%%%%%%%%%%%%%%%%%%%%%%%%%%

\begin{figure*}
\includegraphics[width=1.95\columnwidth]{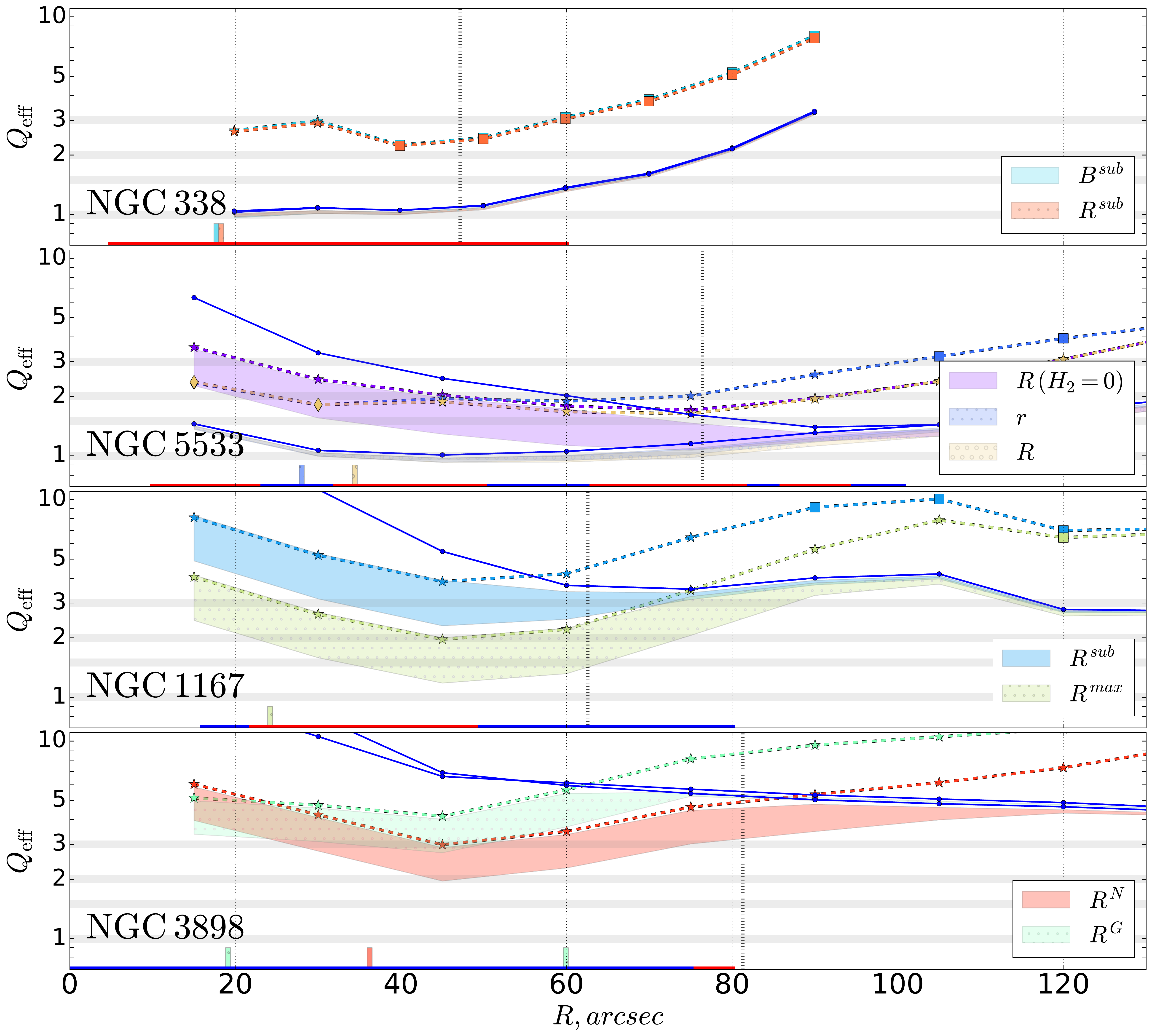}
\caption[width=2.0\columnwidth]{The profile of $Q_\mathrm{eff}$ for examined galaxies. In each figure the solid blue curve shows inverse gas parameter $Q_\mathrm{g}$. Since for some galaxies the model molecular gas distribution depends on the photometry used, figures can contain several such blue curves. For each photometry from Table~\ref{table:photo_parameters} there is corresponding shaded area, which shows possible values of $Q_\mathrm{eff}$ for two-component model. In these areas the upper and lower bounds are determined by the corresponding bounds for $\sigma_R$ from~\eqref{eq:sigR}. Points correspond to distances where gas observational data are available and where we apply criterion of gravitational instability by finding the maximum of Eq.~\eqref{eq:rafikov}. The profile $Q_f$ for a fiducial model is shown by dashed line with different markers which represent the most unstable component (stars for a stellar disc, squares for atomic and diamonds for molecular gas respectively).  Vertical dashed line shows the extension of the line-of-sight stellar velocity dispersion profile $\sigma_\mathrm{los,maj}$. Short vertical segments at the lower axis show disc exponential scale length (or both scale lengths for two-disc models for NGC~3898 and NGC~2985 in Fig.~\ref{fig:allQ2}) for corresponding photometry. The grey semitransparent horizontal lines show the conventional instability thresholds with respect to nonaxisymmetric perturbations. Showed thresholds are $Q_\mathrm{eff} = 1.0;1.5;2.0;3.0$ and corresponding factor values are $\alpha=1.0;0.67;0.5;0.33$. Long horizontal segments at the lower axis indicate regions with large-scale star formation observed in H$\alpha$ line (red), blue areas in SDSS images (blue) or in IR (red dashed in Fig.~\ref{fig:allQ2}). For NGC~5533 there are two models of molecular gas distribution that results in two different blue curves corresponding to the one-fluid criterion of gravitational instability.}
\label{fig:allQ1} 
\end{figure*}

\begin{figure*}
\includegraphics[width=1.95\columnwidth]{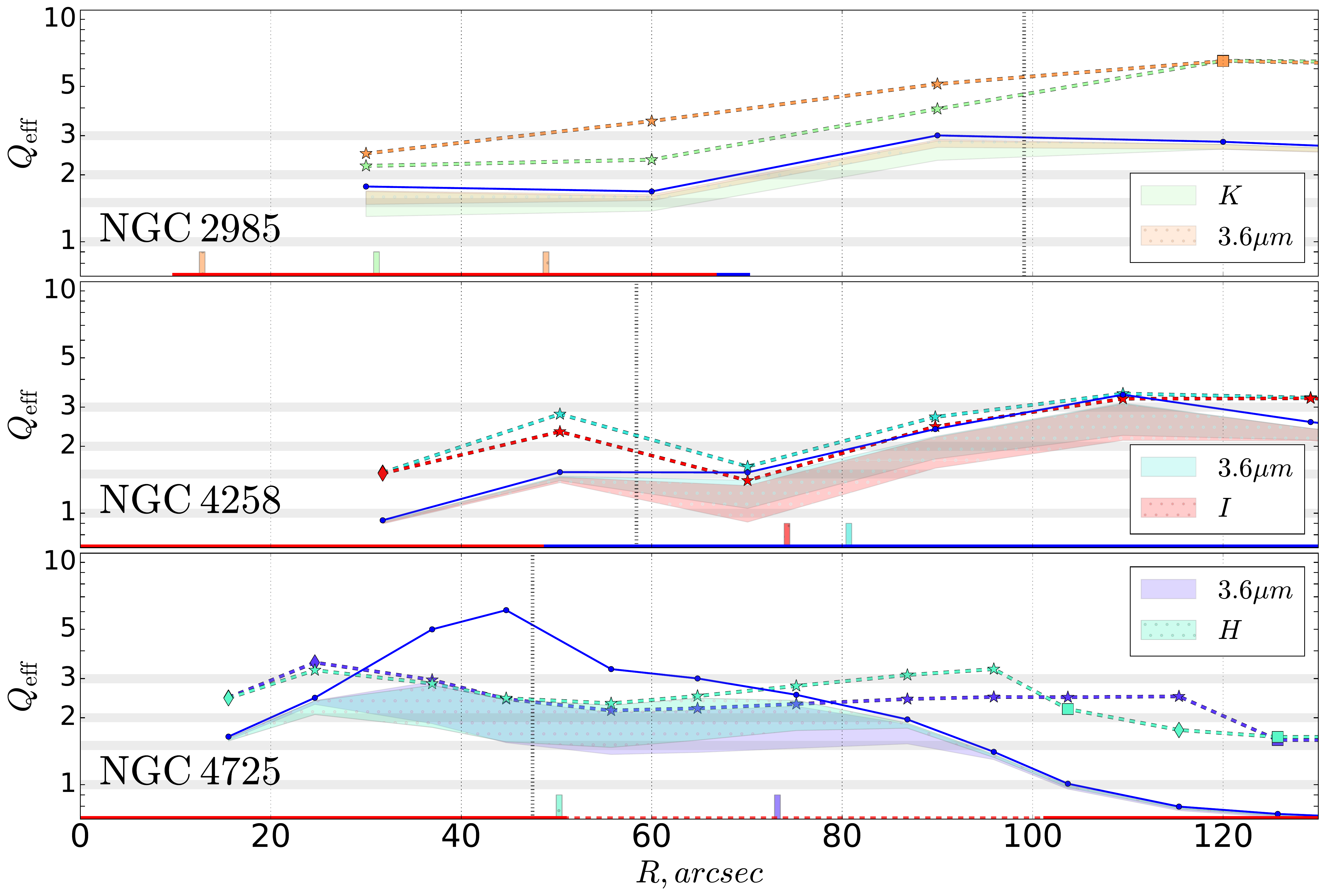}
\caption[width=2.0\columnwidth]{The same as in Fig.~\ref{fig:allQ1} for NGC~2985, NGC~4258 and NGC~4725.}
\label{fig:allQ2} 
\end{figure*}

Figs.~\ref{fig:allQ1}--\ref{fig:allQ2} show the results of applying  two-component instability criterion to the galaxies under study. All galaxies are divided into three groups.

The first group includes NGC~338 and NGC~5533. For them a simple single-fluid criterion gives an unstable disc $Q_\mathrm{g} < 1.5$. In these galaxies star formation is completely determined by a massive gaseous disc (Fig.~\ref{fig:allQ1}). Adding a stellar disc almost does not change $Q_\mathrm{eff}$. Unfortunately the exact profile of $\Sigma_\mathrm{H_2}$ is unknown for both galaxies, so the result may be dependent on the assumed distribution of molecular hydrogen (see below). However, the amount of atomic hydrogen is already enough for a disc to be unstable in the areas of large-scale star formation. Consequently, the influence of stars can be neglected.

The second group is the most interesting. For NGC~1167 and NGC~3898 the value of $Q_\mathrm{g}$ is so large that a gaseous disc remains stable. Despite a large total amount of gas in the disc, it is distributed widely and surface densities  $\Sigma_\mathrm{g}$ are small. At the same time, large rotational velocities point out that stellar discs in this pair of galaxies are very massive, especially in the case of NGC~1167 (Fig.~\ref{fig:data_1167}). Adding such a massive stellar disc changes the dynamic status of the galaxy. $Q_\mathrm{eff}$ decreases noticeably and the star-gas disc becomes unstable $Q_\mathrm{eff}\approx 2-3 $ precisely in the regions of star formation. In these galaxies large-scale star formation is driven by stars. For NGC~1167 the stars impact is stronger because even for the sub-maximal disc its $\Sigma_\mathrm{s}$ are greater and $\Sigma_\mathrm{g}$ are smaller than in NGC~3898. The result for NGC~3898 is contradictory because the SFR is large and numerous blue areas in the disc make it similar to the galaxies from the first group. At the same time the level of instability is less than in NGC~1167, where star formation is weak. Also in NGC~3898 it is possible to see a change in the level of $Q_\mathrm {eff}$ for different photometries. For example, in order to obtain instability for the data from \citet{Gutierrez_etal2011} we need to assume $\alpha \le 1/3 $ for several observable points. Note that model without molecular gas in NGC~5533 is similar to results in this group.

All three remaining galaxies show intermediate results (Fig.~\ref{fig:allQ2}), similar to the first group, but with some features. Note that only these galaxies have an observational profile $\Sigma_\mathrm{H_2}$ instead of a model one.

Among the galaxies in the third group, NGC~4258 and NGC~4725 are similar. Both galaxies possess massive bars and outer spirals with intense star formation. The gas surface densities are large and the gas disc exhibits instability in almost all regions of the observed star formation. The distant outer spirals and the outer ring in NGC~4725 are also unstable according to  simple one-fluid criterion (see Fig.~\ref{fig:outer_spirals}). In general,  simple criterion is sufficient to explan star formation in these galaxies and the level of $Q_\mathrm{g}$ is in excellent agreement with the observed large-scale star formation. As before, adding a star disc makes the model less stable. In case of NGC~4258 this allows to explain an observed star formation at $R>100\arcsec$ as well as radial IR structures $50-100 \arcsec$ for NGC~4725. It should be noted that small values of $Q_\mathrm{eff}$ at the centre of NGC~4258 can be result of the bar influence. 

The last remaining galaxy NGC~2985 shows similar results. The profile of $Q_\mathrm{g}$ consists of three parts: an unstable internal one, where star formation is observed, a marginally stable external area $Q_\mathrm{g} \approx 3$ and an intermediate region between them. The observed star formation regions almost exactly correspond to the instability criterion $Q_\mathrm{g} < 2$. The addition of a stellar disc practically does not affect the dynamic status of the galaxy. Besides for a more accurate two-disc model from the infrared data an decrease in $Q_\mathrm{eff}$ is less than in case of the $K$-band photometry. The impact of the molecular gas on the dynamic status of NGC~2985 is significant. Removing molecular gas from the gas disc makes it stable as in the NGC~3898 while the effective Toomre parameter for the two-component model remain at the same level as before.

%%%%%%%%%%%%%%%%%%%%%%%%%%%%%%%%%%%%%%%%%%%%%%%%%%%%%%%%%%%%%%%%%%%%%
%%%%%%%%%%%%%%%%%%%%%%%%%%%%%%%%%%%%%%%%%%%%%%%%%%%%%%%%%%%%%%%%%%%%%

%%%%%%%%%%%%%%%%%%%%%%%%%%%%%%%%%%%%%%%%%%%%%%%%%%%%%%%%%%%%%%%%%%%%%
%%%%%%%%%%%%%%%%%%%%%%%%%%%%%%%%%%%%%%%%%%%%%%%%%%%%%%%%%%%%%%%%%%%%%

%%%%%%%%%%%%%%%%%%%%%%%%%%%%%%%%%%%%%%%%%%%%%%%%%%%%%%%%%%%%%%%%%%%%%
\subsection{Fiducial model}
%%%%%%%%%%%%%%%%%%%%%%%%%%%%%%%%%%%%%%%%%%%%%%%%%%%%%%%%%%%%%%%%%%%%%

In order to test our assumptions and to see how different our models from usually used ones, we will introduce a simple fiducial model. The model was constructed using Eq.~(19) of \citet{Romeo_Falstad2013}, which is more general that the approximation~\eqref{eq:RW2}. 
We took into account three components (the stellar disc, the atomic and molecular components) as in the application illustrated by \citet{Romeo_Falstad2013}. The reason why we use the approximation rather than the exact solution is that it is widely used and thus can be easy compared with previous results \citep{Westfall_etal2014, Fathi2015, Hallenbeck_etal2016, Garg_Banerjee2017, Romeo_Mogotsi2017}. This approximation also shows a good accuracy and takes into account the disc thickness and different gaseous dispersions effects. We follow \citet{Romeo_Falstad2013} and choose observationally supported gas dispersion values $\sigma_\mathrm{H\,{\sc i}}=11~\mathrm{km\,s^{-1}}$ and $\sigma_\mathrm{H_2}=6~\mathrm{km\,s^{-1}}$. For a stellar disc we adopt our averaged parameters as fiducial. We set the ratio $\sigma_z/\sigma_R=0.5$ and the velocity dispersion equal to a half sum of the used upper and lower dispersion boundaries. For the approximation~\eqref{eq:RW2} this leads to the thickness correction coefficient $T_\mathrm{s}=1.15$ for a stellar disc. Both molecular and atomic discs have isotropic velocity ellipsoid and hence $T=1.5$.

For all galaxies we constructed such averaged and more physically motivated fiducial models and calculated the profiles of $Q_f$. In Figs.~\ref{fig:allQ1}-\ref{fig:allQ2} they are shown by dashed lines with markers. Figures show that $Q_f$ values lie almost everywhere above those found from the two-component model $Q_\mathrm{eff}$, because both thickness and higher atomic gaseous dispersion have significant stabilizing effect. However, there are two different types of profiles depending on $Q_\mathrm{g}$. A fiducial model profile lie within 5-8\% range to the upper $Q_\mathrm{eff}$ limit at the region where $Q_\mathrm{g} > Q_f$. At other radii $Q_f$ become noticeably larger than $Q_\mathrm{g}$ and even more than $Q_\mathrm{eff}$. This is especially clearly seen for NGC~1167 and NGC~4725. There are two reasons behind this behaviour. The first one is in the fact that in the areas where $Q_\mathrm{g} > Q_f$ a stellar component dominates in the local stability level even for thick models. Thus $Q_f$ is close to $T_\mathrm{s}Q_\mathrm{s}$ from Eq.~\eqref{eq:RW2}. The second reason is that a half-sum of the upper and lower $\sigma_R$ limits from~Eq.\eqref{eq:sigR} for inclinations $i=36-64\degr$ is very close to the upper limit if multiplied the by thickness factor $T_\mathrm{s}=1.15$. The difference is less than 8\% for $35\degr < i < 50\degr$ and decreases below 5\% for $i > 50\degr$. Therefore in star dominated regions $Q_f \approx Q_\mathrm{s}$ for the upper $\sigma_R$ limit. For other points with $Q_\mathrm{g} < Q_f$ even if stars still dominate in the stability level, the gaseous terms are no longer negligible. 

We can also notice from the marker types on the fiducial profile in  Figs.~\ref{fig:allQ1}-\ref{fig:allQ2}, that the stellar disc is the most unstable component almost everywhere. There are a few points in the inner parts of galaxies where $\mathrm{H_2}$ is the most unstable component and H\,{\sc i} dominates in distant regions. The one exception is NGC~338 where the disc stability level defined by the atomic hydrogen surface density. The last additional remark is about the dissipation. As we mentioned above the stabilizing effect of the thickness may be compensated by the destabilizing effect of gas dissipation \citep{Elmegreen2011}. It seems like if the dissipation effect is taken into account it will decrease fiducial values of $Q_f$ and they will appear to be inside the founded range of possible two-component models at least in stellar-dominated regions. 

%%%%%%%%%%%%%%%%%%%%%%%%%%%%%%%%%%%%%%%%%%%%%%%%%%%%%%%%%%%%%%%%%%%%%
\subsection{Uncertainties}
%%%%%%%%%%%%%%%%%%%%%%%%%%%%%%%%%%%%%%%%%%%%%%%%%%%%%%%%%%%%%%%%%%%%%

In this paper we use some assumptions that can affect $Q_\mathrm{eff}$ values. The resulting uncertainties for examined models with thin discs are discussed in this subsection.

Errors in galaxy inclination $i$ measurement are first source of uncertainty for $Q_\mathrm{eff}$. For each galaxy we analyzed the scatter of $i$ in different papers and calculate the average angle (see Table~\ref{table:main_parameters}). The greatest error in the inclination angle is equal to $8\degr$, but it does not exceed a couple of degrees for other galaxies. The $i$ value affects how accurately the rotation curve was corrected and thus how large is $\varkappa$. This parameter is also used for finding the constraints on 
$\sigma_R$ and the estimate of $M/L$.
Thus the influence of $i$ on $Q_\mathrm{eff}$ is nonlinear and depends on many factors. For example, one of the most significant changes in $Q_\mathrm{eff}$ for varying $i$ within it limits is for NGC~1167, although the uncertainty error in its inclination is only $2\degr$. But even in this case errors are rather small and not exceed 5\%. Note that $Q_\mathrm{g}$ profile can also change as $i$ changes since it depends on the epicyclic frequency. For all other galaxies the change in $Q_\mathrm{eff}$ while $i$ varies within limits of errors is also small with the exception of NGC~338 and NGC~4725, where in some regions the discrepancy is greater, but the instability level for them remains at the conventional range. Therefore one can neglect the effect of errors in the inclination angle.

Among all parameters that directly participate in the formula for $Q_\mathrm {eff}$ only sound speed $c_\mathrm{g}$ can exert a significant influence, since all other parameters, including $\sigma_R$, are derived directly from the observations. Indeed, observations show \citep{Mogotsi_etal2016,Romeo_Mogotsi2017} that common assumption about $c_\mathrm{g}$ constancy in the galaxy disc as in \citet{Kennicutt1989,Leroy_etal2008} can be incorrect. 
The value of sound speed 6~km\,s$^{-1}$ adopted here is rather small and therefore may be inconsistent with the real  values. \citet{Romeo_Mogotsi2017} show that $c_\mathrm{g}$ value for molecular and atomic gas in central regions to be around 50-100~km\,s$^{-1}$, but in areas under interest sound speed falls within the range 4-20~km\,s$^{-1}$. In order to check how the variation of $c_\mathrm{g}$ affects the effective Toomre parameter we divided this range uniformly into 50 bins. For each bin the profile $Q_\mathrm{eff}$ was calculated separately and then the average profile and the standard deviation were determined.

\begin{figure*}
\includegraphics[width=1.7\columnwidth]{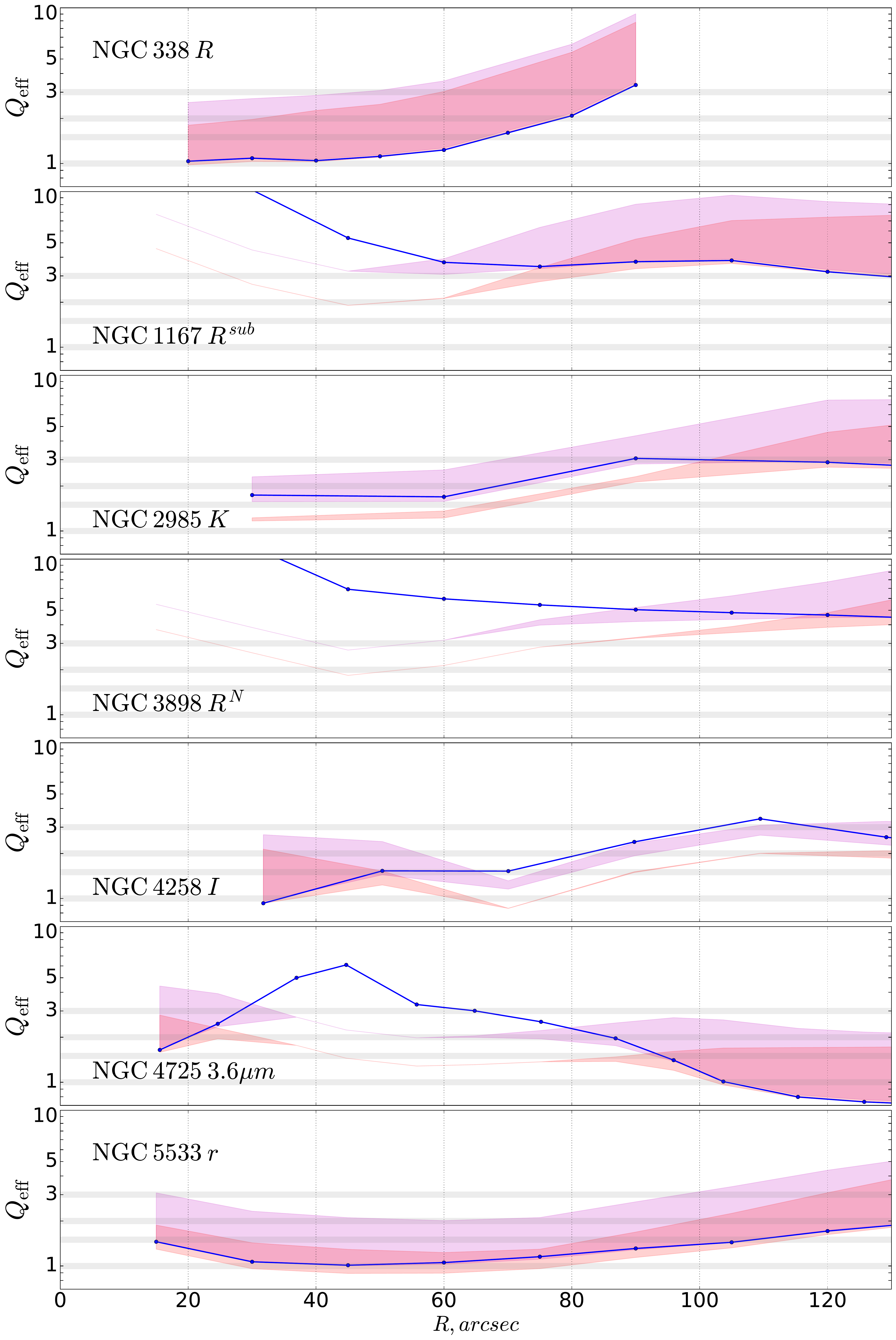}
\caption{$Q_\mathrm{eff}$ change for different $c_\mathrm{g}$ values. The solid blue line shows $Q_\mathrm{g}$ profile for $c_\mathrm{g}=6$~km\,s$^{-1}$ as before. Filled areas show how upper (orange) and lower (magenta) boundaries of $Q_\mathrm{eff}$ scatter will change if $c_\mathrm{g}$ varies for a photometry given in the upper left corner. These areas show the mean boundary level with an error of one standard deviation calculated for 50 models where $c_\mathrm{g}$ vary from 4~km\,s$^{-1}$ to 20~km\,s$^{-1}$. It can be seen that the change is small and the difference from the model with $c_\mathrm{g}=6$~km\,s$^{-1}$ is minimal if the initial level of instability was contributed by stars ($Q_\mathrm{g} > Q_\mathrm{eff}$).}
\label{fig:cg_depend}
\end{figure*}

The results are shown in Fig.~\ref{fig:cg_depend}. The curves with dots correspond to the one-fluid level of $Q_\mathrm{g}$ for $c_\mathrm{g}= 6$~km\,s$^{-1}$ as before and the changes in the upper and lower bounds of the two-component models are shown by filled areas. In many regions  $Q_\mathrm{eff}$ lie above than $Q_\mathrm{g}$ like in NGC~338, since most tested models are calculated for sound speed greater than 6~km\,s$^{-1}$. It is clearly seen that the shape of the profiles is similar to those in Figs.~\ref{fig:allQ1}--\ref{fig:allQ2} and the biggest uncertainties can be seen in regions where gas component is more unstable (the second peak in Fig.~\ref{fig:338_vs_3898_Qeff}). This is due to the fact that sound speed $c_\mathrm{g}$ in Eq.~\eqref{eq:rafikov} is included in the second term only, which corresponds to the gas contribution to instability. Therefore as Fig.~\ref{fig:cg_depend} shows, the uncertainty grows in regions where two-component models were close to one-fluid model with $c_\mathrm{g} = 6$~km\,s$^{-1}$. Comparison with Figs.~\ref{fig:allQ1}--\ref{fig:allQ2} shows that uncertainty in $Q_\mathrm{eff}$ related to the thickness effect and hot atomic gas in the fiducial model is almost equal to that produced by the variation of $c_\mathrm{g}$. As expected, variation of $c_\mathrm{g}$ changes models for NGC~338, NGC~5533 and for outer regions of NGC~4725 the most of all, but even for them the disc still remains mostly unstable. In all other cases the impact of $c_\mathrm{g}$ on the dynamic status of a  galaxy stays small and does not affect the result. This conclusion is especially valid for NGC~1167 and NGC~3898, which almost do not show changes in $Q_\mathrm{eff}$ in star-forming regions (Fig.~\ref{fig:cg_depend}). It should be noted that the real change may be larger in the inner regions and less in the outer areas, since used sound speed limits 4-20~km\,s$^{-1}$ seems to be smaller and greater than real values for these regions, respectively.

%%%%%%%%%%%%%%%%%%%%%%%%%%%%%%%%%%%%%%%%%%%%%%%%%%%%%%%%%%%%%%%%%%%%%
%%%%%%%%%%%%%%%%%%%%%%%%%%%%%%%%%%%%%%%%%%%%%%%%%%%%%%%%%%%%%%%%%%%%%

We also tested the models for galaxies with modelled $\Sigma_\mathrm{H_2}$ profiles. We increase the scale length of gas distribution from $h$ to $2h$: $\Sigma_\mathrm{H_2} \propto \exp(-R/2h)$, where $h$ is the stellar disc scale length. In other words we redistributed the same amount of gas in a wider area. The largest effect up to 25\% was for NGC~338. As expected, central regions of the galaxy became more stable, whereas for distant areas the values of $Q_\mathrm{eff} $ decrease. For NGC~1167 and NGC~3898 the molecular gas redistribution does not change the results even in central regions.

%%%%%%%%%%%%%%%%%%%%%%%%%%%%%%%%%%%%%%%%%%%%%%%%%%%%%%%%%%%%%%%%%%%%%
\subsection{$Q_\mathrm{eff}$ approximations}
%%%%%%%%%%%%%%%%%%%%%%%%%%%%%%%%%%%%%%%%%%%%%%%%%%%%%%%%%%%%%%%%%%%%%

As a rule, in papers concerning the analysis of two-component instability an approximate formula is used instead of the exact solution of Eq.~\eqref{eq:rafikov}. We compared the results obtained from the direct solution with the estimates from a simple approximation by \citet{Wang_Silk1994} as well as with a more accurate formula by \citet{Romeo_Wiegert2011}. For each observational point in radii range $R<130\arcsec$ and for each accepted photometry we calculated approximate values $Q_\mathrm{WS}$ using Eq.~\eqref{eq:WS} and $Q_\mathrm{RW}$ using Eq.~\eqref{eq:RW1}. The adopted sound speed is 6~km\,s$^{-1}$ as before. The total number of points in comparison is twice as observational ones  since for each distance we calculated $Q_\mathrm{eff}$ and approximations for the upper and lower bounds of $\sigma_R$. 

%%%%%%%%%%%%%%%%%%%%%%%%%%%%%%%%%%%%%%%%%%%%%%%%%%%%%%%%%%%%%%%%%%%%%
\begin{figure*}
\includegraphics[width=2\columnwidth]{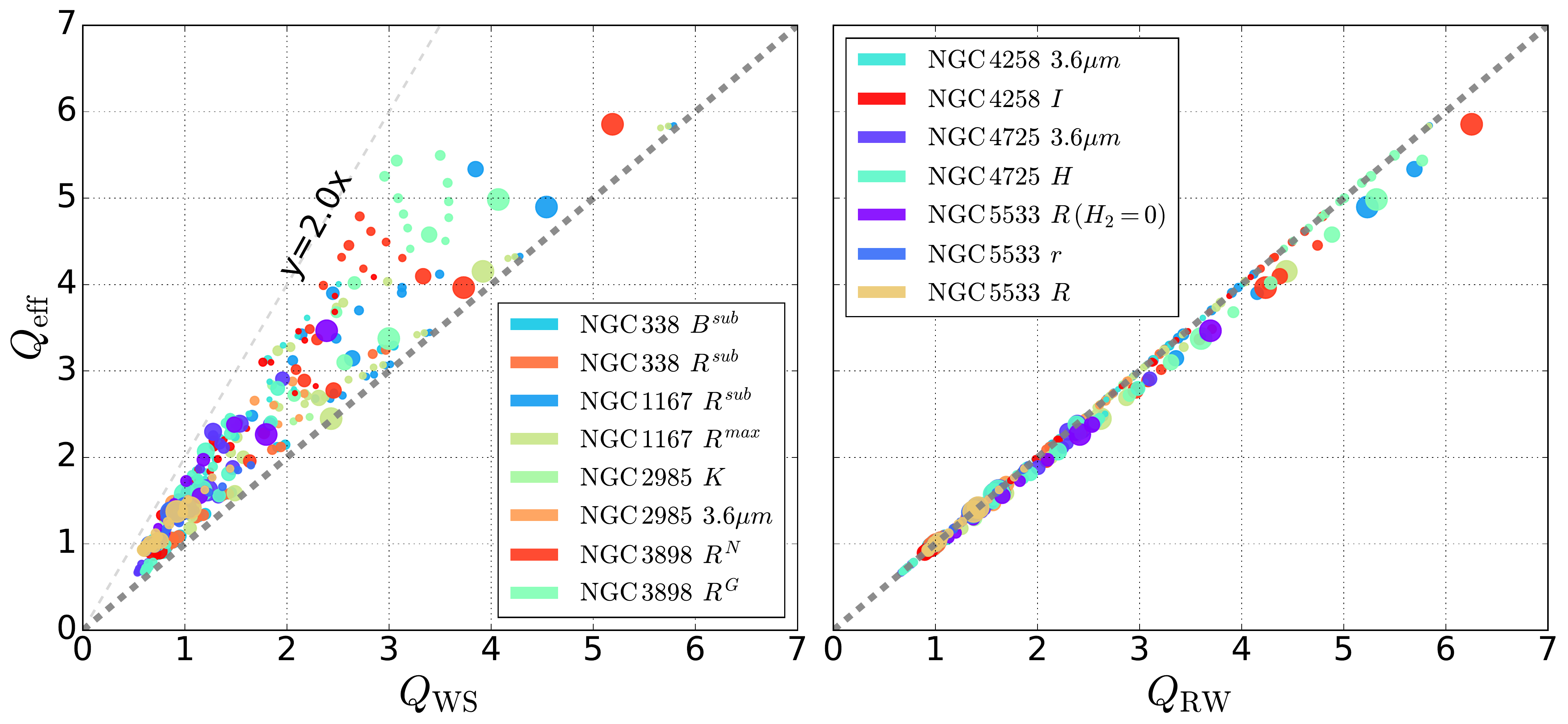}
\caption{The effective parameter $Q_\mathrm{eff}$ found from the Eq.~\eqref{eq:rafikov} versus $Q_\mathrm{WS}$ obtained from  \citet{Wang_Silk1994} (left) and $Q_\mathrm{RW}$ from  \citet{Romeo_Wiegert2011} (right) approximations. Each point corresponds to the observation point from Figs.~\ref{fig:allQ1}--\ref{fig:allQ2} taken either for the lower or upper bounds of $\sigma_R$. The point size reflects the distance from the centre, the smaller the point, the farther it is from the centre. The dashed line is the line where $Q_\mathrm{eff}$ is equal to  $Q_\mathrm{WS}$ or $Q_\mathrm{RW}$ in the left and right plots, respectively. Legend is the same for both plots and splitted between them.
}
\label{fig:WS_RF} 
\end{figure*}
%%%%%%%%%%%%%%%%%%%%%%%%%%%%%%%%%%%%%%%%%%%%%%%%%%%%%%%%%%%%%%%%%%%%%

Results of comparison are shown in Fig.~\ref{fig:WS_RF}. It is clearly seen that the approximation~\eqref{eq:WS} is rather inaccurate and on the average it gives the error in $Q_\mathrm{eff}$ around 30\%, but in extreme cases can decrease effective Toomre parameter up to twice from the non approximated value. Even so it is still used in recent works \citep{Stark2017, Fisher2017, Krumholz2016}. The disc in this approximation is estimated to be more unstable since both dependent on $\bar{k}$ multipliers in Eq.~\eqref{eq:2fluidJS1} are always less than 1. In contrast, the formula~\eqref{eq:RW1} from \citet{Romeo_Wiegert2011} leads to the root mean square error around 3\% and the maximum difference is around 7\%, which agrees with the conclusion in the original paper. Using of $Q_\mathrm{RW}$ makes two-component disc more stable than in the exact solution, but only within a few percent.

It is remarkable that observational points in the right plot of Fig.~\ref{fig:WS_RF} fall into two lines $Q_\mathrm{RW} \approx Q_\mathrm{eff}$ and $Q_\mathrm{RW} \approx 1.07\times Q_\mathrm{eff}$. This is due to the form of the approximation formula~\eqref{eq:RW1}, where weights are determined by the more unstable component. First line corresponds to the case when a gaseous disc is more unstable and second line appears when a stellar disc is more unstable. The factor 1.07 can be explained as follows. If the stellar disc is more unstable, the maximum $Q_\mathrm{eff}^{-1}$ falls at the first peak in Fig.~\ref{fig:338_vs_3898_Qeff}, which is always located near the point $\bar{k}\approx 1$ as the maximum of the function $\displaystyle \frac{1-\exp(-x^2)\,I_0(x^2)}{x}$. The corresponding weights in $Q_\mathrm{RW}^{-1}$ for this case are  $W_\mathrm{s} = 1$, $W_\mathrm{g} = \displaystyle \frac{2s}{1+s^2}$. Hence the second term in Eq.~\eqref{eq:rafikov} coincides with that in $Q_\mathrm{RW}^{-1}$ and can be neglected since it is small in comparison with the first term. Therefore in the case of unstable stellar disc we have:  $$\frac{Q_\mathrm{RW}}{Q_\mathrm{eff}}\approx2\times\frac{1-\exp(-1)I_0(1)}{1}\approx 1.07.$$

%%%%%%%%%%%%%%%%%%%%%%%%%%%%%%%%%%%%%%%%%%%%%%%%%%%%%%%%%%%%%%%%%%%%%
\subsection{Instability levels}
%%%%%%%%%%%%%%%%%%%%%%%%%%%%%%%%%%%%%%%%%%%%%%%%%%%%%%%%%%%%%%%%%%%%%

The gravitational instability criterion for nonaxisymmetric perturbations contains a factor $\alpha$, which for the one-fluid case range from $\approx1/3$ \citep{Hunter_etal1998} to $\approx1/2$ \citep{Kennicutt1989,Martin_Kennicutt2001}. The corresponding star formation threshold $Q_\mathrm{g} < 2-3$ also follows from the theoretical \citep{Morozov1985, Griv2012} and numerical \citep{Li_etal2005} investigations. Also there are some evidences of even a higher level of instability ($Q_\mathrm {g} < 2-4$ in  \citealp{Zasov_Zaitseva2017} and $Q_\mathrm{RW} \approx 1-4$ in star-forming spirals from THINGS, see figure 5 in \citealp{Romeo_Wiegert2011}). However there is still no general consensus about the exact instability threshold.

\begin{figure}
{%
  \includegraphics[width=\columnwidth]{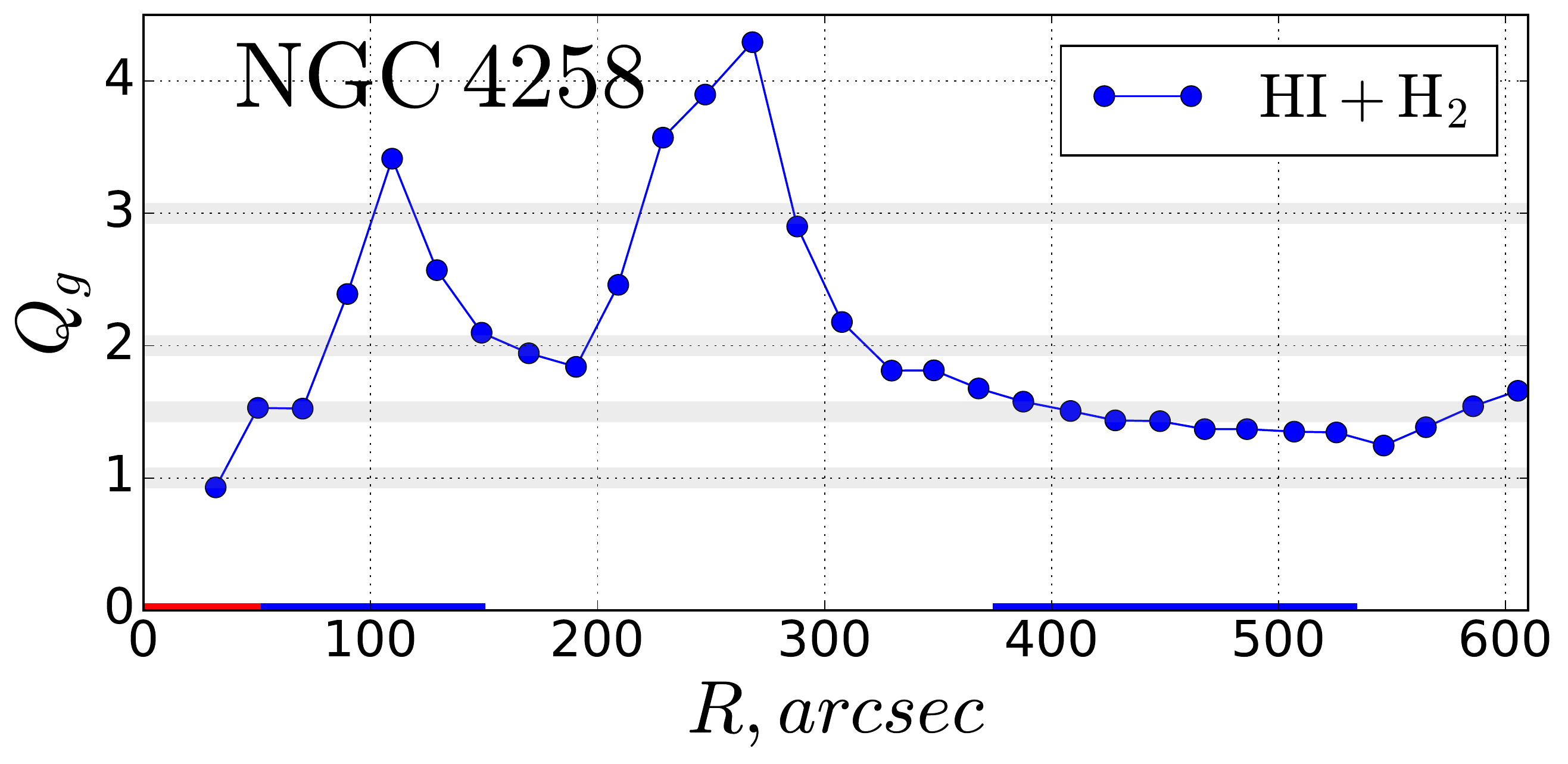}%
}
{%
  \includegraphics[width=\columnwidth]{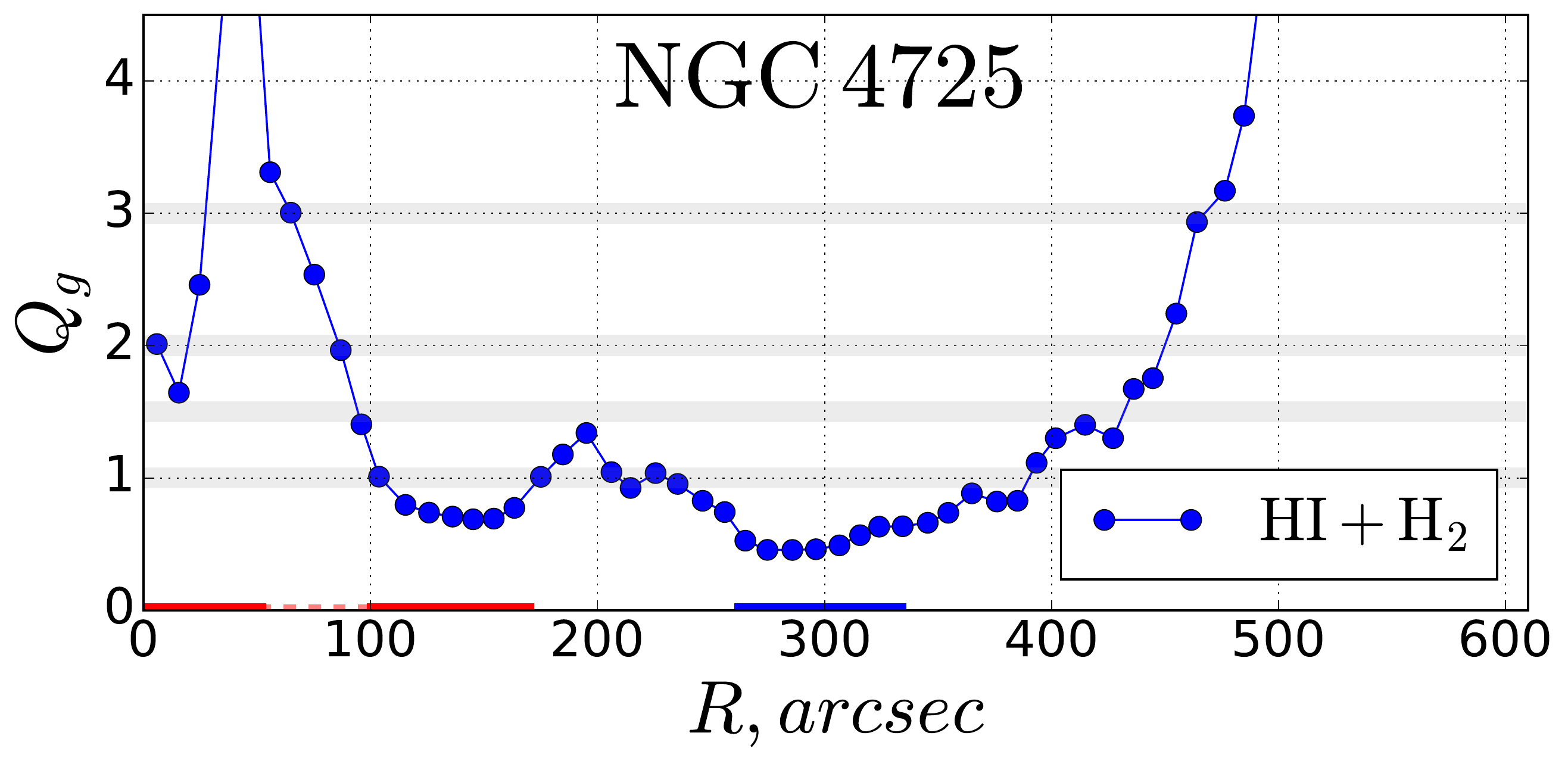}%
}
\caption{The profile $Q_\mathrm{g}$ in outer regions of NGC~4258 and NGC~4725. The notations are similar to those in Fig.~\ref{fig:allQ1}. The gaseous disc is unstable in outer spirals where large-scale star formation is observed.}
\label{fig:outer_spirals}
\end{figure}

Values of $Q_\mathrm{g}$ in Figs.~\ref{fig:allQ1}--\ref{fig:allQ2} for all galaxies except two  are in agreement with the generally accepted $\alpha$. The one-fluid criterion shows instability in regions with noticeable  star formation rate and does not demonstrate instability in regions without it. Central parts of galaxies with large gas surface densities (NGC~338 and NGC~5533) as well as external spirals and the outer ring for NGC~4258 and NGC~4725 (see Fig.~\ref{fig:outer_spirals}) are unstable even for $\alpha \approx 2/3$, which is close to the value originally proposed in \citet{Kennicutt1989}. For NGC~2985 the areas where $Q_\mathrm{g} < 2$ ($\alpha = 0.5$) agrees well with regions of star formation. At more remote areas gaseous disc of NGC~2985 becomes marginally stable on level $Q_\mathrm{g} \approx 3$ as well as disc in NGC~1167.

%%%%%%%%%%%%%%%%%%%%%%%%%%%%%%%%%%%%%%%%%%%%%%%%%%%%%%%%%%%%%%%%%%%%%
%%%%%%%%%%%%%%%%%%%%%%%%%%%%%%%%%%%%%%%%%%%%%%%%%%%%%%%%%%%%%%%%%%%%%

However, two galaxies have gaseous discs which are significantly stable according to the one-fluid criterion. To a greater extent this is correct for NGC~3898 with $Q_\mathrm {g} > 5 $ and to a less degree for NGC~1167 with $Q_\mathrm {g} > 4-5$ in star-forming regions. For these two galaxies, the destabilizing effect of the stellar disc becomes crucial. The model of maximal stellar disc for NGC~1167 allows to explain star formation and gives $\alpha \approx 0.5$ whereas the sub-maximal model leads to $\alpha \approx 0.33$ or more. As for NGC~3898, in the case of more massive model of a stellar disc~\citet{Noordermeer_vanderHulst2007} value of $\alpha$ that is sufficient  for explanation approximately equal to 1/3 and for the second model it is too small, approximately 1/4. Except this last model, all others give the correction factor for nonaxisymmetric perturbations in agreement with generally accepted values. Notice also that all fiducial models except NGC~3898 and a sub-maximal model for NGC~1167 demonstrate $Q_f < 3$ in star-forming regions.

In all other cases the destabilizing effect of the stellar disc decreases $Q_\mathrm{eff}$, but rather slightly and without any impact on $\alpha$ value. Note that one possible exception is a model without molecular gas for NGC~5533. One-fluid criterion for atomic gas can not explain the observed star formation in central regions while two-component criterion can. But we can expect a lot of molecular gas in the centre, and this gas can be sufficient to explain star formation. Another interesting feature for two-component model is observed in NGC~4725, where $Q_\mathrm{eff}$ gives the unstable disc with $\alpha \approx 0.5$ in the radial star-forming region at $50-100 \arcsec$. However, in this area the influence of the bar is great, which is difficult to take into account. And we can not estimate how the azimuthal averaging of the gas surface density in the bar area affects the final result.

The azimuthal averaging of data in strongly asymmetric galaxies is also a factor that affects $\alpha$ and makes a galaxy more stable. The value of $\alpha$ can change up to two times \citep{Martin_Kennicutt2001}. However, this could lead to instability in one-fluid model for NGC~3898 and NGC~1167 only and they have well-defined symmetrical structure.

The underestimation of both the molecular gas and sound speed $c_\mathrm{g}$ also affects $\alpha$, but only in central regions where the amount of molecular gas and $c_\mathrm{g}$ can be large \citep{Mogotsi_etal2016,Romeo_Mogotsi2017}. However in this paper we consider more distant areas beyond the bulge. We tested what would happen if the molecular gas will be redistributed more widely or assuming larger $c_\mathrm{g}$. Fig.~\ref{fig:cg_depend} show that $\alpha$ can change substantially, but nonetheless has remained higher than $1/3$. 

Rejection of the assumption that stellar radial velocity dispersions continued at the level of the last observation point can only increase $\alpha$ since real $\sigma_R$ values at distant regions are likely to be less than used ones. Maximal disc model, which significantly affects $Q_\mathrm{eff}$, is justified for those galaxies that are very bright and fast rotating, hence they have heavy stellar disc, especially NGC~1167. Neglecting the disc thickness and gas dissipation is consistent with the obtained values of $\alpha$ \citep{Elmegreen2011} as already mentioned. Finally, there are other types of instability criteria and laws which can explain the observed large-scale star formation. Most of them described in \citet{Leroy_etal2008, Kennicutt2012}. We did not examine them here, since star formation in analyzed galaxies is consistent with one-fluid or two-component gravitational instability criteria and the correction factor $\alpha$ is similar to the values obtained earlier.

%%%%%%%%%%%%%%%%%%%%%%%%%%%%%%%%%%%%%%%%%%%%%%%%%%%%%%%%%%%%%%%%%%%%%
\section{Conclusions}
%%%%%%%%%%%%%%%%%%%%%%%%%%%%%%%%%%%%%%%%%%%%%%%%%%%%%%%%%%%%%%%%%%%%%

The explanation of the observed large-scale star formation is important for studying the evolution of spiral galaxies and physical processes, which take place in their discs. The most frequently used model explains the formation of new stars as a consequence of gaseous  discs fragmentation due to gravitational instability. Simple one-fluid criterion of the instability is well-consisted with observations if a galaxy posses a large amount of gas and its surface densities are large \citep{Kennicutt1989}, but it shows the stable disc in other cases. For such galaxies one should consider a modified criterion, the so-called two-component instability criterion \citep{Jog_Solomon1984, Rafikov2001}.

In this paper we applied a two-component criterion to observational data along the major axis for 7 spiral galaxies of early types. There are several significant differences in the performed analyses comparing to similar previous works. First of all, in order to obtain stellar velocity dispersion values in the radial direction $\sigma_R$ we used extended line-of-sight velocity dispersion profiles along the major axis $\sigma_\mathrm{los,maj}$. Profiles of $\sigma_R$ were constrained using dynamic considerations instead of using a constant ratio $\sigma_z/\sigma_R$ or an averaged thickness relation \citet{Kregel_etal2002}, as in most papers. It is important to take into account $\sigma_R$ scatter, since the radial velocity dispersion of stars becomes one of the main sources of errors. We used the maximal and sub-maximal discs as mass models of stellar disc instead of popular photometric calibration relations in the form by \citet{Bell_etal2003}, because such relations gave strongly unmatched models for photometries in different bands. We have taken into account the effect of possible variation of sound speed $c_\mathrm{g}$. For all galaxies we analyzed gravitational instability in extended regions, up to distances $R \le 130 \arcsec$, or up to several exponential disc scale lengths. Finally, the values of the effective Toomre parameter $Q_\mathrm{eff}$ were obtained directly from the exact dispersion relation without using the approximate solutions, whose accuracy we also checked.

As a result of our analysis we came to the following conclusions:

--- The two-component criterion differs slightly from the one-fluid in the case when galactic disc contains a large amount of gas. Then a simple one-fluid criterion is sufficient to explain observed star formation as for NGC~338.

--- The two-component criterion can explain the large-scale star formation in those regions where a simple criterion gives stable gaseous disc. This is significant for  galaxies NGC~1167 and NGC~3898.

--- The case of NGC~1167 is outstanding in our sample because the galaxy has a very heavy stellar disc with total mass greater than $10^{11}M_{\sun}$. Even in the case of sub-maximal model this heavy disc destabilizes a gaseous disc, which was stable before. This result does not depend on used assumptions about gas parameters. Thus NGC~1167 is an example of a spiral galaxy in which star formation is entirely driven by stars. Similar result was obtained for NGC~1068 in \citet{Romeo_Fathi2016}, but using an approximate solution.

--- The galaxy NGC~3898 demonstrates a marginal stability of combined star-gas disc while it shows much higher rate of star formation than NGC~1167. Upon that, gaseous disc has large margin of gravitational stability. Probably, in NGC~3898 the role of nonaxisymmetric perturbations is great or star formation is regulated by another instability mechanism.

--- For all galaxies, gravitational instability criteria can explain observed large-scale star formation if we take into account nonaxisymmetric perturbations. Performed analysis implies that the instability threshold for the two-component case is $Q_\mathrm{eff} < 1.5-2.5$ (except NGC~3898, for which $Q_\mathrm{eff} < 3$). Such a threshold agrees with previous results obtained in both observational \citep{Kennicutt1989,Hunter_etal1998,Martin_Kennicutt2001} and theoretical \citep{Morozov1985,Li_etal2005} papers.

--- Taking into account the thickness and higher H\,{\sc i} dispersion effects in a three-component fiducial model from \citet{Romeo_Falstad2013} makes discs more stable than in the used two-component models. However in areas of interest where the stellar component almost entirely determines the dynamic status of a galaxy, the fiducial model becomes close to the upper limit of the two-component model and the dissipation effect can lower it even more.

--- It is checked that for our sample the use of the approximate formulas can lead to errors in $Q_\mathrm{eff}$ estimation up to 50\% for \citet{Wang_Silk1994} approximation and up to 7\% for \citet{Romeo_Wiegert2011} approximation if gaseous disc is more stable than the stellar one.

%%%%%%%%%%%%%%%%%%%%%%%%%%%%%%%%%%%%%%%%%%%%%%%%%%%%%%%%%%%%%%%%%%%%%%
\section*{Acknowledgements}
%%%%%%%%%%%%%%%%%%%%%%%%%%%%%%%%%%%%%%%%%%%%%%%%%%%%%%%%%%%%%%%%%%%%%

We are grateful to Alexei Moiseev and Ivan Katkov for providing spectral data. We thank Alessandro Romeo for his great review and highly appreciate the comments and suggestions that significantly contributed to improving the quality of the article.

This research makes use of the NASA/IPAC Extragalactic
Database (NED) which is operated by the Jet Propulsion Laboratory, California Institute of Technology, under contract with the National Aeronautics and Space Administration, the LEDA database (\href{url}{http://leda.univ-lyon1.fr}) and images from Sloan Digital Sky Survey. Funding for the Sloan Digital Sky Survey has been provided by the Alfred P. Sloan Foundation, the U.S. Department of Energy Office of Science, and the Participating Institutions. SDSS acknowledges support and resources from the centre for High-Performance Computing at the University of Utah. The SDSS web site is \href{url}{www.sdss.org}.
%%%%%%%%%%%%%%%%%%%%%%%%%%%%%%%%%%%%%%%%%%%%%%%%%%%%%%%%%%%%%%%%%%%%%%

%%%%%%%%%%%%%%%%%%%%%%%%%%%%%%%%%%%%%%%%%%%%%%%%%%%%%%%%%%%%%%%%%%%%%
\appendix
%%%%%%%%%%%%%%%%%%%%%%%%%%%%%%%%%%%%%%%%%%%%%%%%%%%%%%%%%%%%%%%%%%%%%
\section{Individual galaxies}
%%%%%%%%%%%%%%%%%%%%%%%%%%%%%%%%%%%%%%%%%%%%%%%%%%%%%%%%%%%%%%%%%%%%%

%%%%%%%%%%%%%%%%%%%%%%%%%%%%%%%%%%%%%%%%%%%%%%%%%%%%%%%%%%%%%%%%%%%%%
${\bf NGC~338}$ is a galaxy of the Sa or Sab type highly inclined ($i=64\degr$). It is described in detail in \citet{Zasov_etal2012}. Distances estimated to be in the range 65-75~Mpc. Bright spirals in the disc and dust lane in the centre can be seen.

The galaxy is highly asymmetric as it can be seen from the rotation curve, where the approaching and receding parts can differ by more than 50~km\,s$^{-1}$. Such asymmetry can be caused by interaction with the UGC 623 satellite. The averaged rotation velocity profile $v_\mathrm{c}$ gradually decreases after the maximum around $15\arcsec$. The velocity dispersion profile along the major axis stretches to $50\arcsec$, where it reaches the plateau. The errors of $\sigma_\mathrm{los, maj}$ are small and its value in the centre is greater than 150~km\,s$^{-1}$.

%%%%%%%%%%%%%%%%%%%%%%%%%%%%%%%%%%%%%%%%%%%%%%%%%%%%%%%%%%%%%%%%%%%%%
\begin{figure*}
\includegraphics[width=1.6\columnwidth]{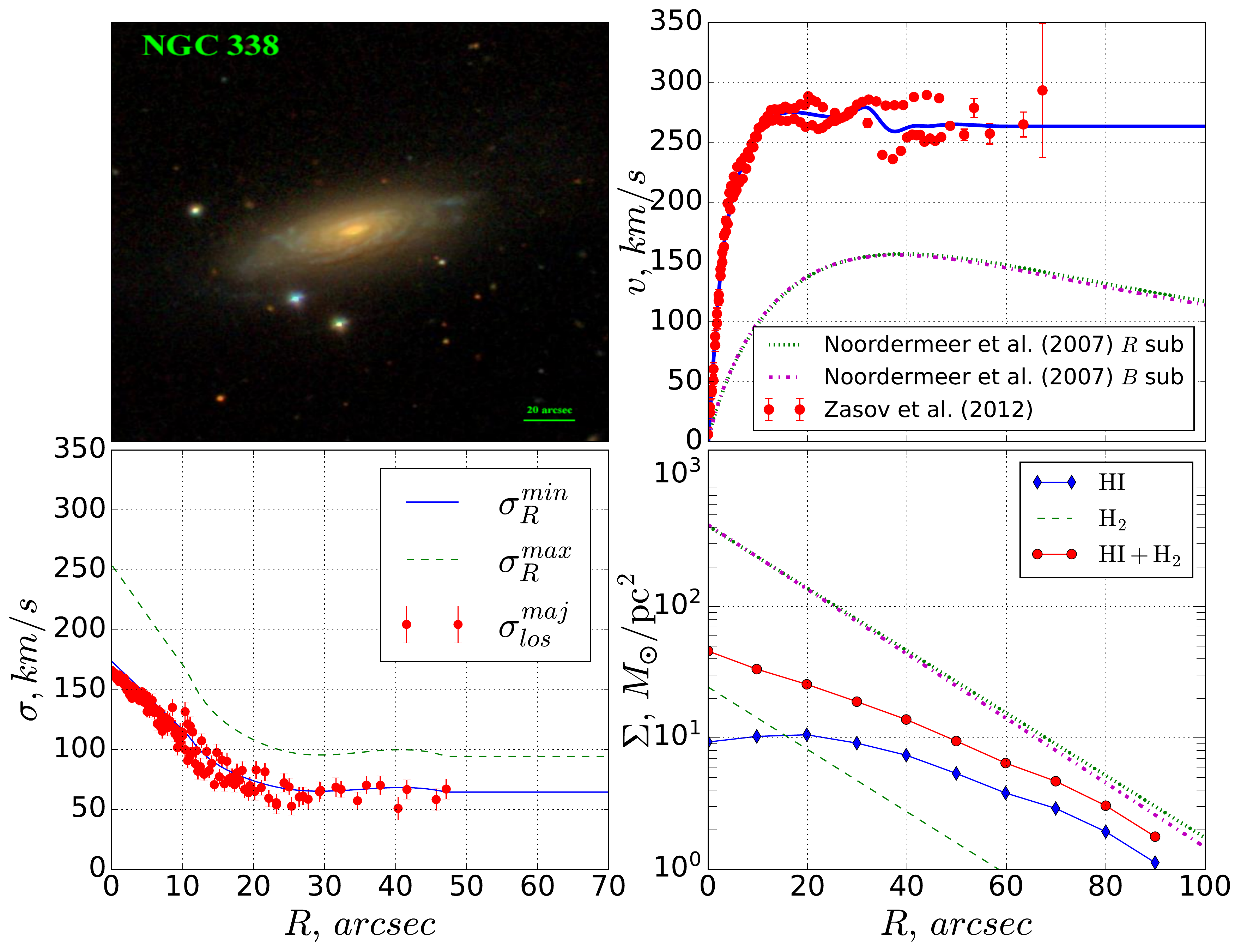}
\caption{Observational data for galaxy NGC~338. The left upper plot shows an image from SDSS and green horizontal line shows the scale. The upper right plot presents the rotation curve. Points show observational data, solid line is a used $v_\mathrm{c}$ approximation and the dashed curves are stellar disc rotation curves corresponding to the used photometry. The bottom left plot shows the observed velocity dispersion profile along the major axis $\sigma_\mathrm{los,maj}$ indicated by points. The upper dashed line and the lower solid line show bounds~\eqref{eq:sigR} of the radial velocity dispersion profile $\sigma_R$. The bottom right plot demonstrates surface density profiles. Solid line with diamonds represents atomic hydrogen, dashed line represents a modelled profile of molecular hydrogen and solid line with circles represents gas total density corrected for the presence of helium. Stellar surface density profiles for used photometries are shown in the same way as in the upper right plot.
}
\label{fig:data_338} 
\end{figure*}
%%%%%%%%%%%%%%%%%%%%%%%%%%%%%%%%%%%%%%%%%%%%%%%%%%%%%%%%%%%%%%%%%%%%%

Asymmetry is also seen in the distribution of H\,{\sc i} \citep{Noordermeer_etal2005}. The total amount of atomic hydrogen in NGC~338 exceeds $10^{10}M_{\sun}$. The largest surface density $\Sigma_{\mathrm{H\, {\sc i}}}$ value is around 11\,$M_{\sun}$\,pc$^{-2}$. The regions of star formation are seen along the spiral arms like blue knots and observed up to the optical boundary of the galaxy at $60\arcsec$ from its centre.

%%%%%%%%%%%%%%%%%%%%%%%%%%%%%%%%%%%%%%%%%%%%%%%%%%%%%%%%%%%%%%%%%%%%%
${\bf NGC~1167}$ is a very bright galaxy of the early type S0, inclined to $36\degr$ \citep{Noordermeer_etal2005}. There are very weak and thin spirals in its disc. The photometric profile is well divided into a disc and a bulge. The spectrum of NGC~1167 shows broad emission lines in the centre and the galaxy is classified as Seyfert galaxy.

The rotation curve stretches to $200\arcsec$, which is equal to around 8 exponential disc scale lengths. The values of the rotational velocity are very large and at the maximum reach 377~km\,s$^{-1}$ at the distance $110\arcsec$ from the centre \citep{Struve_etal2010}. The velocity dispersion profile along the major axis  $\sigma_\mathrm{los,maj}$ extends up to $60\arcsec$ \citep{Zasov_etal2008} and shows large central values around 200~km\,s$^{-1}$. This profile is in good agreement with CALIFA map \citep{CALIFA_2017}.

%%%%%%%%%%%%%%%%%%%%%%%%%%%%%%%%%%%%%%%%%%%%%%%%%%%%%%%%%%%%%%%%%%%%%
\begin{figure*}
\includegraphics[width=1.6\columnwidth]{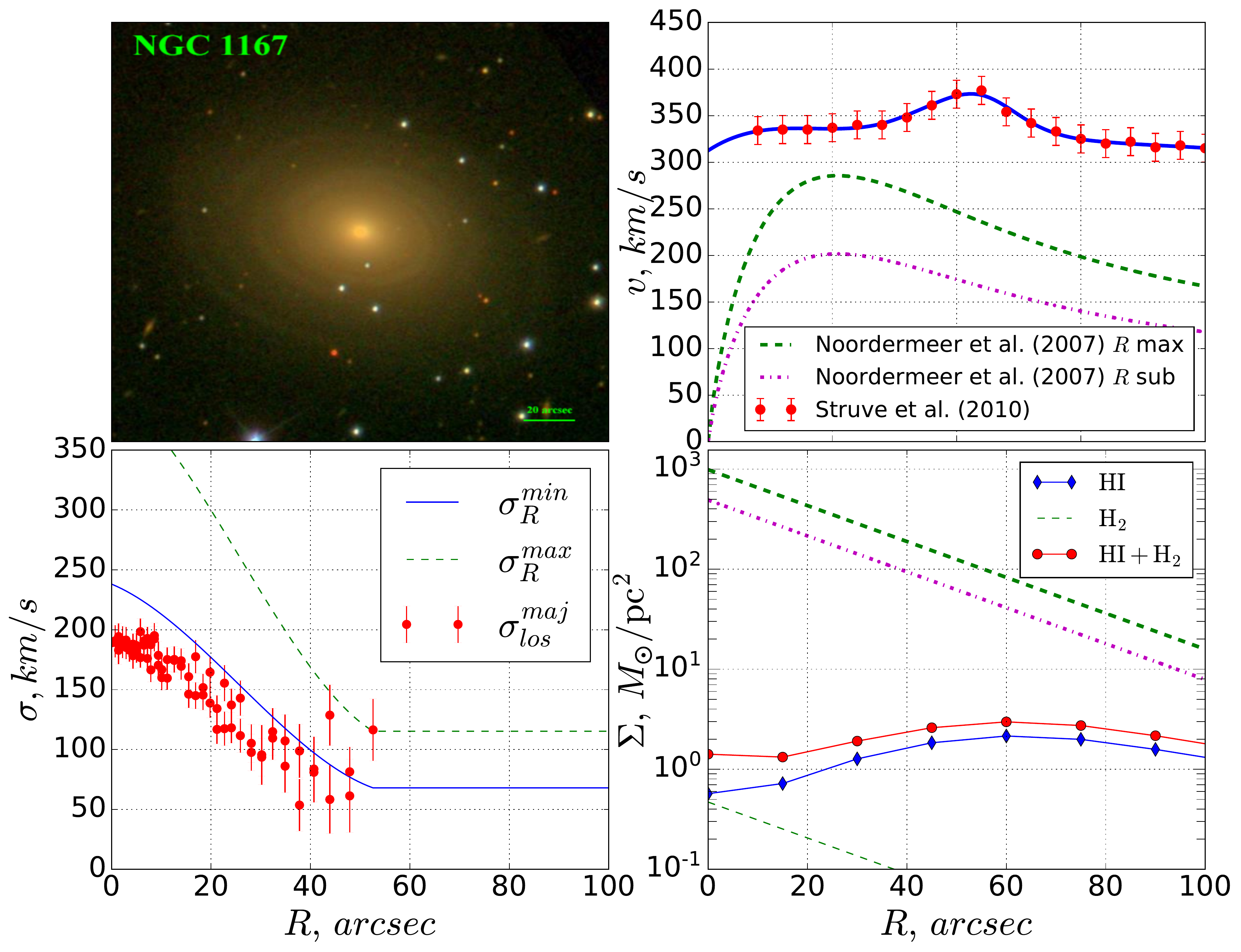}
\caption{The same as in Fig.~\ref{fig:data_338} for NGC~1167.}
\label{fig:data_1167} 
\end{figure*}
%%%%%%%%%%%%%%%%%%%%%%%%%%%%%%%%%%%%%%%%%%%%%%%%%%%%%%%%%%%%%%%%%%%%%

The $\Sigma_{\mathrm{H\, {\sc i}}}$ profile from \citet{Noordermeer_etal2005,Struve_etal2010} demonstrates maximum density values around 2\,$M_{\sun}$\,pc$^{-2}$ near the gas ring at $60\arcsec$ from the centre. The total amount of the atomic gas in the galaxy is quite large and exceeds $10^{10} M_{\sun}$. The data about the molecular hydrogen were found in \citet{OSullivan_etal2015}. The authors noted, that it is not likely that some CO emission outside of the 7.8 kpc beam was missed. The total amount of $\mathrm{H_2}$ is around ten times less than the mass of the atomic gas.

There are no noticeable blue areas in the SDSS images of NGC~1167. \citet{CALIFA_2016} used unsharp-masking technique which allowed them to measure and to map the spiral-like star-forming pattern with surface brightness $\mu_r\simeq 25^\mathrm{m}/\sq\arcsec$ and extension up to $40\arcsec$ from the centre. They also added  the well-agreed H$\alpha$ contours and justified that such star formation was not a consequence of a merging event.

%%%%%%%%%%%%%%%%%%%%%%%%%%%%%%%%%%%%%%%%%%%%%%%%%%%%%%%%%%%%%%%%%%%%%
${\bf NGC~2985}$ is a regular grand design spiral galaxy of the early Sab type at the intermediate inclination of $37\degr$ \citep{Noordermeer_etal2008}. In the outer part of the galaxy a massive spiral arm encircles the galaxy. This spiral arm forms a pseudo ring near $70\arcsec$ which is revealed as a ``hump'' at the photometric profile \citep{Noordermeer_vanderHulst2007,Gutierrez_etal2011}. NGC~2985 is not included in SDSS and hence Digitized Sky Survey (DSS) and Hubble Space Telescope (HST) images were used. According to the AINUR survey \citep{Comeron_etal2010} the galaxy has a central gas ring with a radius around 50~pc.

%%%%%%%%%%%%%%%%%%%%%%%%%%%%%%%%%%%%%%%%%%%%%%%%%%%%%%%%%%%%%%%%%%%%%
\begin{figure*}
\includegraphics[width=1.6\columnwidth]{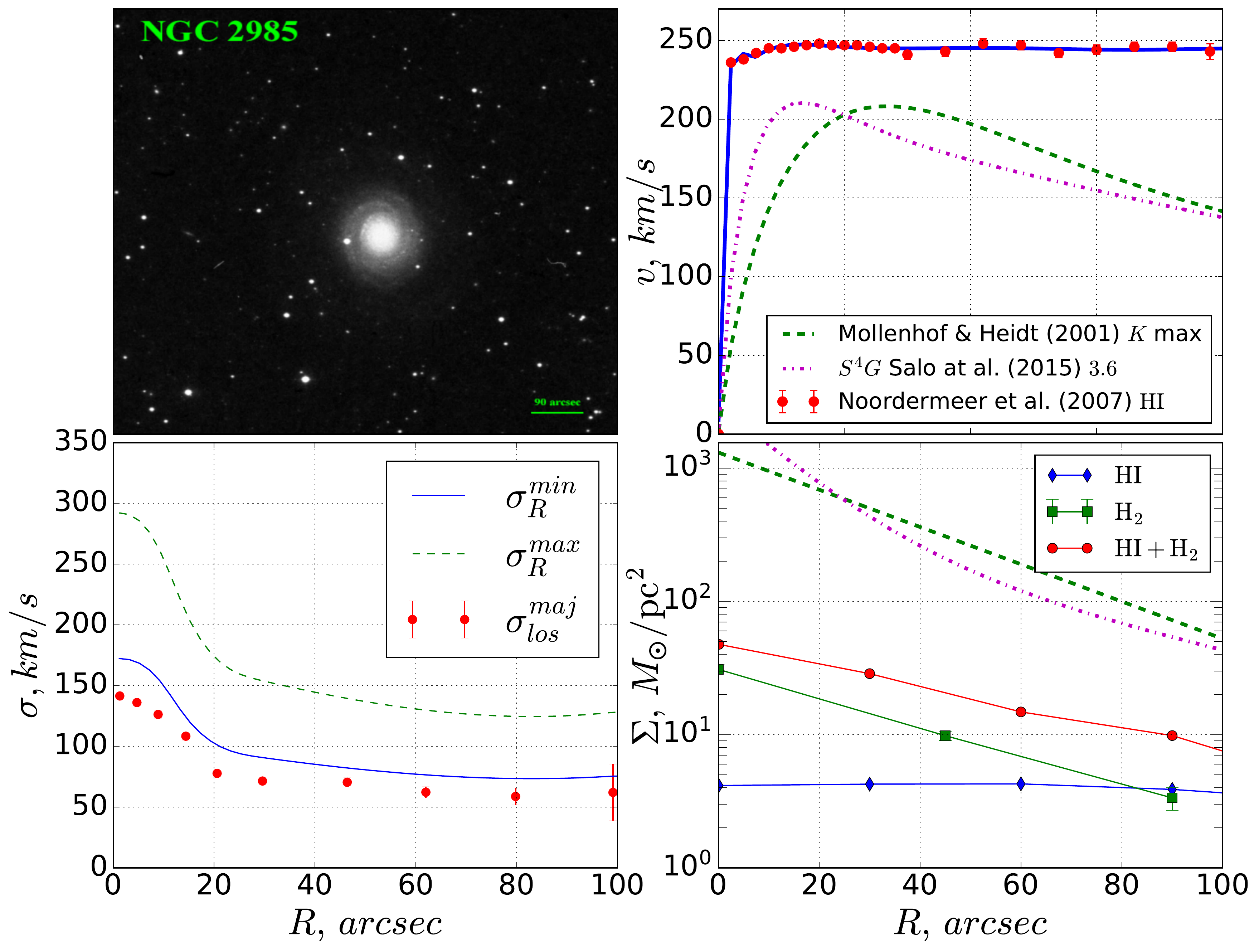}
\caption{The same as in Fig.~\ref{fig:data_338} for NGC~2985.  The left upper plot shows an image from DSS. In the bottom right plot the solid curve with squares represents the observed surface densities of molecular hydrogen. The profile $\Sigma_\mathrm{H_2}$ extends up to distance $90\arcsec$, after which the surface density $\mathrm{H_2}$ is assumed to be zero.}
\label{fig:data_2985} 
\end{figure*}
%%%%%%%%%%%%%%%%%%%%%%%%%%%%%%%%%%%%%%%%%%%%%%%%%%%%%%%%%%%%%%%%%%%%%

The velocity dispersion profiles along the major axis were found in \citet{Heraudeau_etal1999,Noordermeer_etal2008,Gerssen_etal2000}. Data from \citet{Gerssen_etal2000} has large uncertainties. All bibliographic profiles $\sigma_\mathrm{los, maj}$ are consistent with each other, except one peculiarity near $R\approx 25\arcsec$. The profile in \citet{Noordermeer_etal2008} is very long with farthest observation point located around $100\arcsec$ from the centre of the galaxy. The rotation curve is almost constant up to $250\arcsec$ with velocities equal to 240~km\,s$^{-1}$.

A large number of isolated blue knots are present throughout the galactic disc as it can be seen from HST and H$\alpha$ \citep{Hameed_Devereux2005} images. The atomic gas is distributed uniformly with the surface density around $4\,M_{\sun}$\,pc$^{-2}$ which remains constant up to the outer spiral boundary and then gradually decreases.

%%%%%%%%%%%%%%%%%%%%%%%%%%%%%%%%%%%%%%%%%%%%%%%%%%%%%%%%%%%%%%%%%%%%%
${\bf NGC~3898}$ has a morphological type SA(s)ab. The inclination angle is rather large, its  estimates varies within $50\degr-70\degr$. The disc contains weak tightly twisted spirals. 

%%%%%%%%%%%%%%%%%%%%%%%%%%%%%%%%%%%%%%%%%%%%%%%%%%%%%%%%%%%%%%%%%%%%%
\begin{figure*}
\includegraphics[width=1.6\columnwidth]{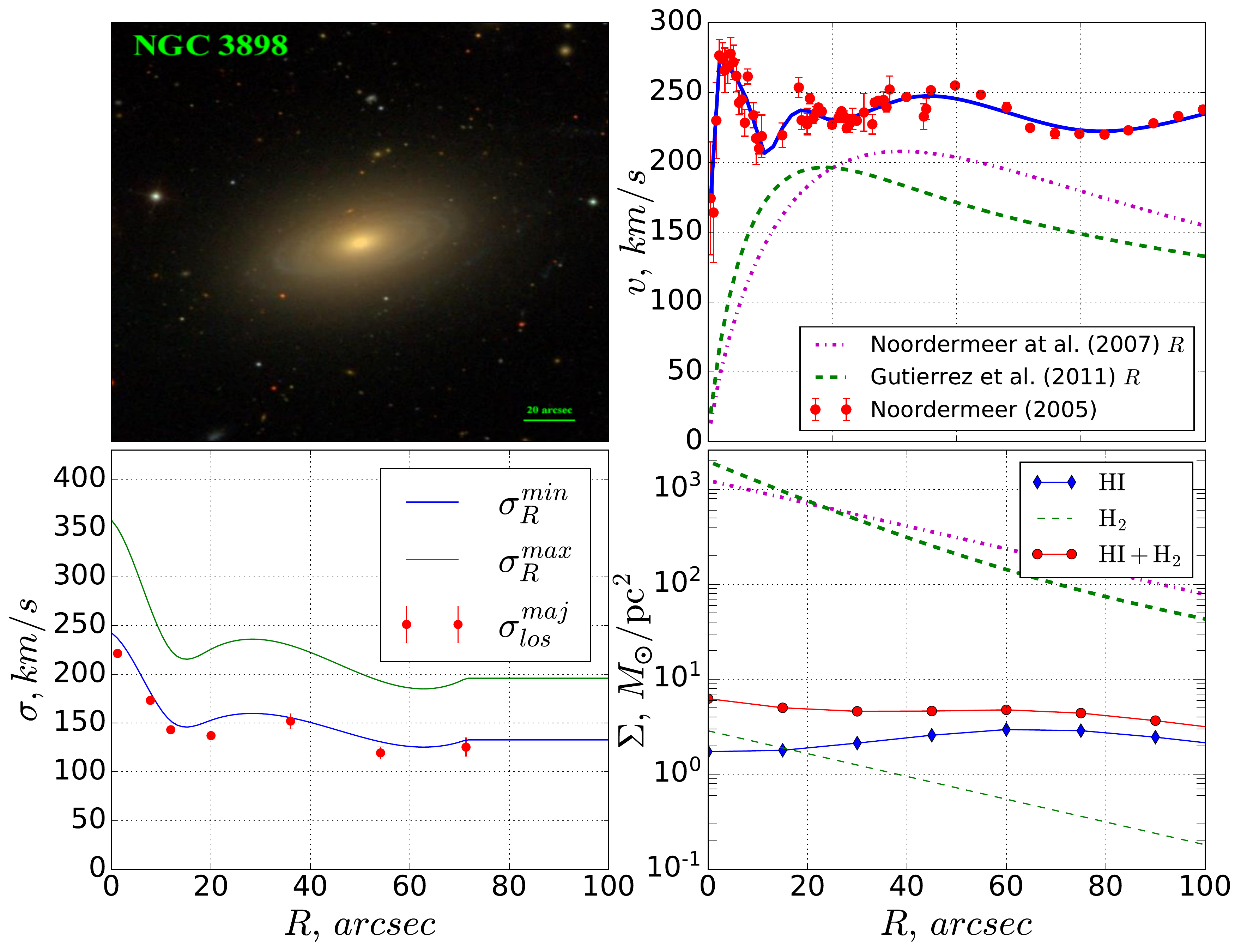}
\caption{The same as in Fig.~\ref{fig:data_338} for NGC~3898.}
\label{fig:data_3898} 
\end{figure*}
%%%%%%%%%%%%%%%%%%%%%%%%%%%%%%%%%%%%%%%%%%%%%%%%%%%%%%%%%%%%%%%%%%%%%

The line-of-sight velocity dispersion profiles along the major axis $\sigma_\mathrm{los, maj}$ are from \citet{Pignatelli_etal2001,Noordermeer_etal2008, Heraudeau_etal1999}. They are consistent with each other except three points in \citet{Heraudeau_etal1999}. They show large central values around 250~km\,s$^{-1}$. The longest profile from \citet{Noordermeer_etal2008} extends up to $70\arcsec$. The rotation curve demonstrates a local maximum near $100\arcsec$ from the centre and a local minimum near $150\arcsec$. After that velocities return to the previous level of 250~km\,s$^{-1}$.

The surface densities of the atomic hydrogen $\Sigma_\mathrm{H\,{\sc i}}$ reach the maximum value of about $3\,M_{\sun}$\,pc$^{-2}$ near $60\arcsec$ and remain constant for next $100\arcsec$ at a half of maximal level \citep{Noordermeer_etal2005}. The total mass of molecular hydrogen in galaxy is 15-20 times less than the total atomic gas  amount. Star formation in NGC~3898 is traced well by H$\alpha$ emission \citep{Hameed_Devereux2005,Pignatelli_etal2001}. In the H$\alpha$ image many densely located knots of new stars are seen up to $70\arcsec$ as well as inside the distant outer spiral.

%%%%%%%%%%%%%%%%%%%%%%%%%%%%%%%%%%%%%%%%%%%%%%%%%%%%%%%%%%%%%%%%%%%%%
${\bf NGC~4258}$ is galaxy of morphological type SABb, inclined to $60\degr-70\degr$. The galaxy demonstrates a large number of features in its structure which include a massive bar, the dust lane and the misaligned inner gas disc \citep{Miyoshi_etal1995,Herrnstein_etal1999}. The central region of the galaxy within $200\arcsec$ contains two tightly twisted spirals, which are sometimes classified as part of a bulge \citep{Fisher_Drory2010}. The bulge is difficult to separate from the disc and sometimes it is classified as pseudobulge \citep{Fisher_Drory2010}. The outer disc is represented by two asymmetric spirals which are very extended. According to \citet{deVaucouleurs_etal1991} $R_{25} \approx 9 $~arcmin. NGC~4258 is also well-known by its anomalously extended arms which are visible in X-ray emission and deviate from the disc plane.

%%%%%%%%%%%%%%%%%%%%%%%%%%%%%%%%%%%%%%%%%%%%%%%%%%%%%%%%%%%%%%%%%%%%%
\begin{figure*}
\includegraphics[width=1.6\columnwidth]{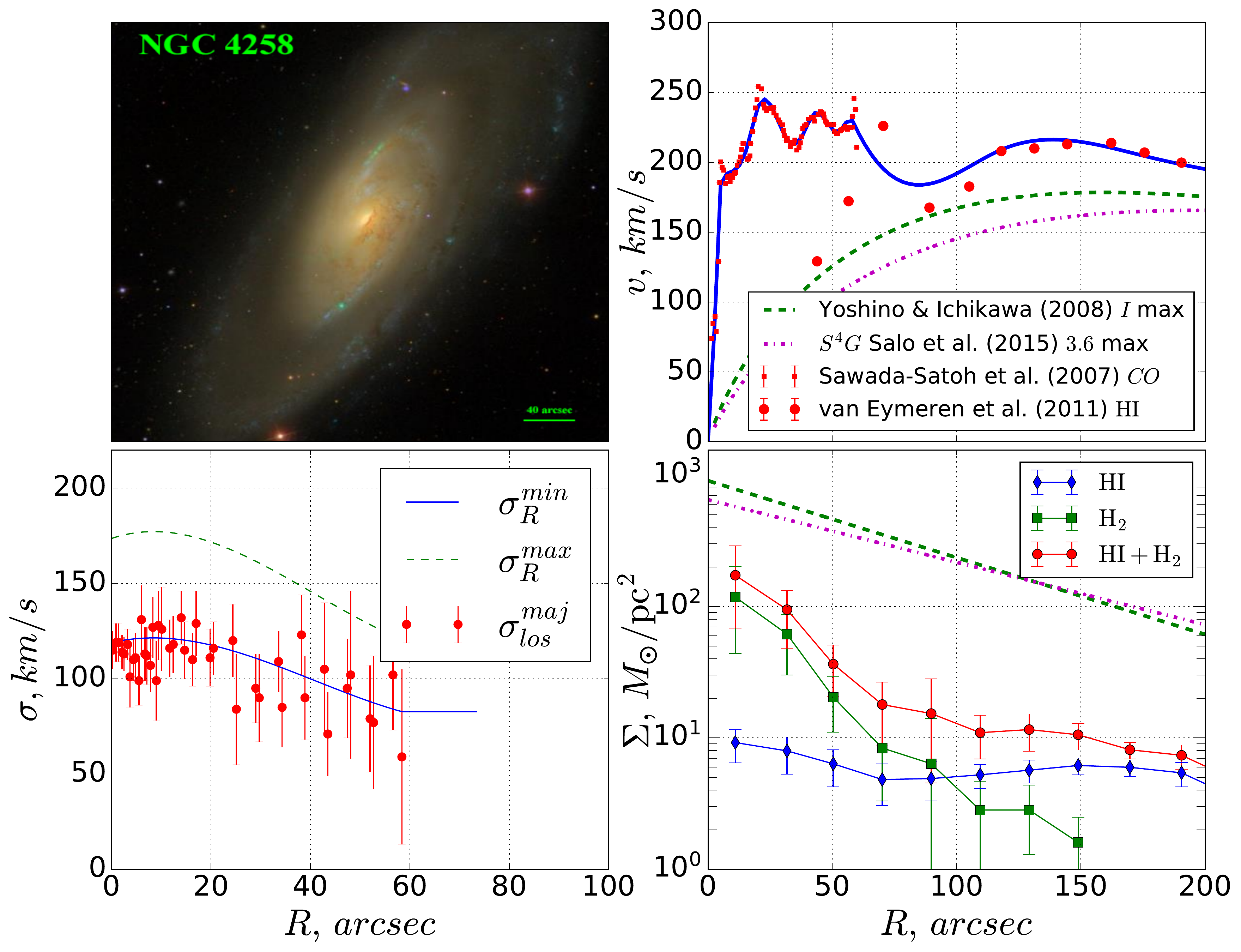}
\caption{The same as in Fig.~\ref{fig:data_338} for NGC~4258. In the bottom right plot a solid curve with squares represents the observed surface densities of molecular hydrogen.}
\label{fig:data_4258} 
\end{figure*}
%%%%%%%%%%%%%%%%%%%%%%%%%%%%%%%%%%%%%%%%%%%%%%%%%%%%%%%%%%%%%%%%%%%%%

The velocity dispersion profile $\sigma_\mathrm{los, maj}$ from \citet{Heraudeau_etal1998} changes a little within 80-120~km\,s$^{-1}$ and extends up to $50\arcsec$. The H\,{\sc i} rotation curve from \citet{Eymeren_etal2011} traces only outer regions at distances greater than $100\arcsec$. Therefore CO curve from \citet{Sawada_Satoh_etal2007} was used to analyze the internal disc. The combined rotation curve decreases after its maximum up to $250\arcsec$ and then starts to increase again.

The galaxy contains a large amount of gas \citep{Yim_etal2016}. The central value of the $\mathrm{H_2}$ surface density obtained from the BIMA survey exceeds $100\,M_{\sun}$\,pc$^{-2}$ and remains above $10\,M_{\sun}$\,pc$^{-2}$ up to $60\arcsec$. The central value of $\Sigma_\mathrm{H\,\sc i}$ is also very large (about $10\,M_{\sun}$\,pc$^{-2}$) and remains constant within $300-500\arcsec$. In the inner and outer spirals blue regions of star formation are visible. Also, the inner area as a whole and outer spirals are visible in FUV and NUV as GALEX data shows \citep{Thilker_etal2007}. The inner region also can be resolved in H$\alpha$ line \citep{Courtes_etal1993} and contains extended irregular structures.

%%%%%%%%%%%%%%%%%%%%%%%%%%%%%%%%%%%%%%%%%%%%%%%%%%%%%%%%%%%%%%%%%%%%%
${\bf NGC~4725}$ is a close Sb/SBb galaxy with a classical bulge, inclined to $44\degr$ \citep{Eymeren_etal2011,Yim_etal2016}. Its internal structure is complex. The galaxy contains a massive bar, oriented almost along the major axis. From the ends of the bar start two spiral arms, which immediately form a ring-like structure with a strong star formation. Outside the ring the spiral structure becomes strongly asymmetric and continues mainly eastward in form of a wide arm.

The velocity dispersion profile $\sigma_\mathrm{los,maj}$ varies a little and extends up to $50\arcsec$. Gas rotation curves from \citet{Noordermeer_etal2005,Eymeren_etal2011} are consistent with each other and represent a smooth curve with a maximum near $120\arcsec$. There are two usually used distance estimations that differ by a factor of two. The Cepheid distance to NGC~4725 was determined to be 13~Mpc \citep{Gibson_etal1999} and it is used in \citet{Fisher_Drory2010,Mollenhoff_Heidt2001} and HYPERLEDA database. In \citet{Eymeren_etal2011, Yim_etal2016} the measured distance is 26~Mpc adopted from the NED database.

%%%%%%%%%%%%%%%%%%%%%%%%%%%%%%%%%%%%%%%%%%%%%%%%%%%%%%%%%%%%%%%%%%%%%
\begin{figure*}
\includegraphics[width=1.6\columnwidth]{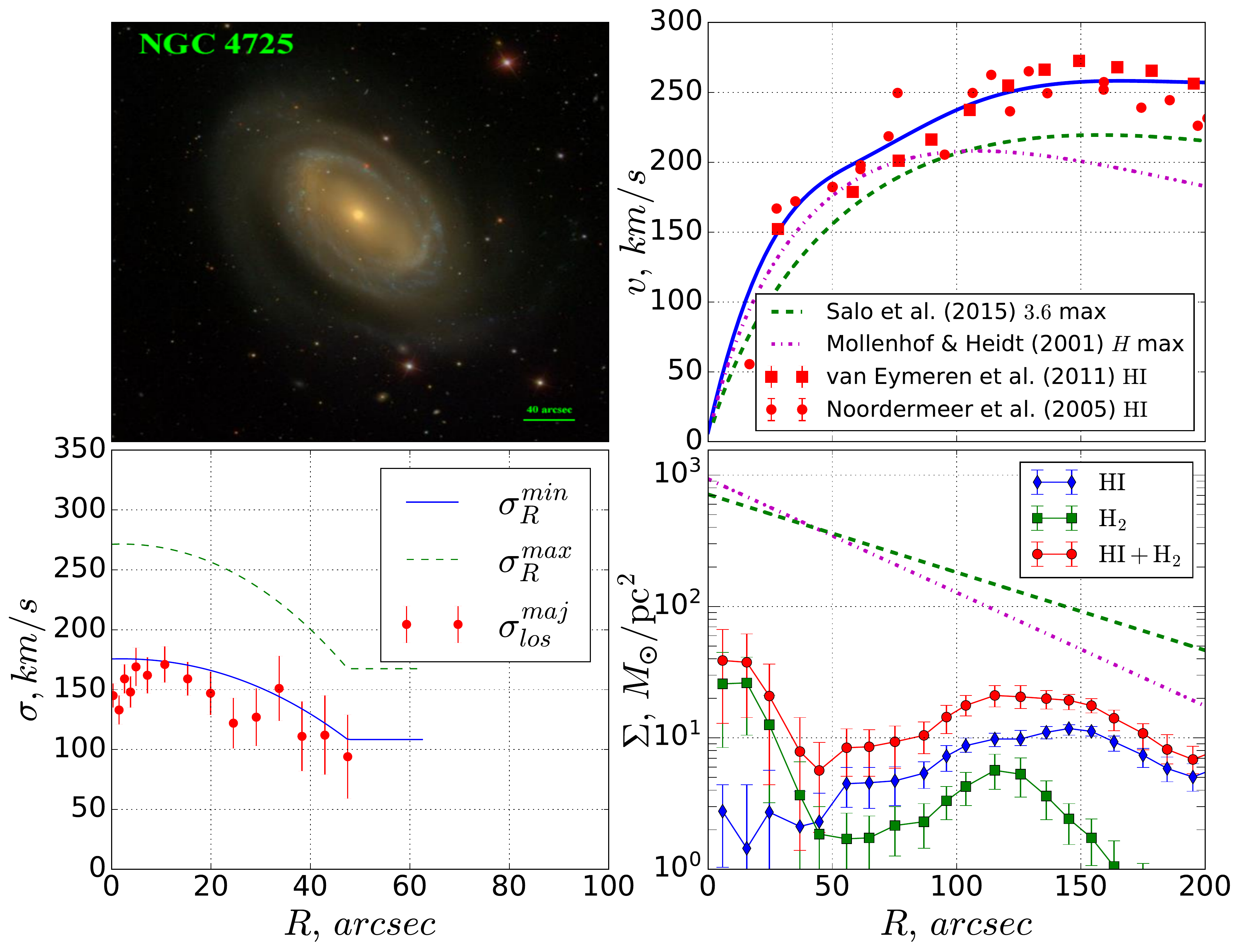}
\caption{The same as in Fig.~\ref{fig:data_338} but for NGC~4725. In the bottom right plot a solid curve with squares represents observed surface densities of molecular hydrogen.}
\label{fig:data_4725} 
\end{figure*}
%%%%%%%%%%%%%%%%%%%%%%%%%%%%%%%%%%%%%%%%%%%%%%%%%%%%%%%%%%%%%%%%%%%%%

Molecular hydrogen is almost not observed inside the central part with the bar and follows the spiral arms in the outer regions. The $\Sigma_\mathrm{H\,\sc i}$ profile from \citet{Noordermeer_etal2005} has two peaks, which correspond to the inner ring and the outer spiral. The second peak is very asymmetrical. The surface density values at the peaks are 4\,$M_{\sun}$\,pc$^{-2}$ and 3\,$M_{\sun}$\,pc$^{-2}$, respectively. The profile of $\Sigma_\mathrm{H\,\sc i}$ from \citet{Yim_etal2016} is also demonstrate two peaks and covers additional regions $<50\arcsec$ and $> 360\arcsec$. This work also provides the $\Sigma_\mathrm{H_2}$ profile, obtained from $\mathrm{CO}(J=2\rightarrow 1)$ observations. It extends up to $200\arcsec$. The surface density of the molecular hydrogen in the centre is 12\,$M_{\sun}$\,pc$^{-2}$ and then rapidly decreases with local maximum at the gas ring at 15~kpc. The total hydrogen mass from \citet{Yim_etal2016} is larger than in \citet{Noordermeer_etal2005} because of significant difference in distance estimations.

H$\alpha$ emission traces the star forming regions and is observed the entire ring and in the outer spiral \citep{Hameed_Devereux2005}. An IR image from SPITZER also shows\footnote{http://www.spitzer.caltech.edu/images/2355-sig05-011-NGC-4725} a noticeable star formation inside the whole bar.

%%%%%%%%%%%%%%%%%%%%%%%%%%%%%%%%%%%%%%%%%%%%%%%%%%%%%%%%%%%%%%%%%%%%%
\begin{figure*}
\includegraphics[width=1.6\columnwidth]{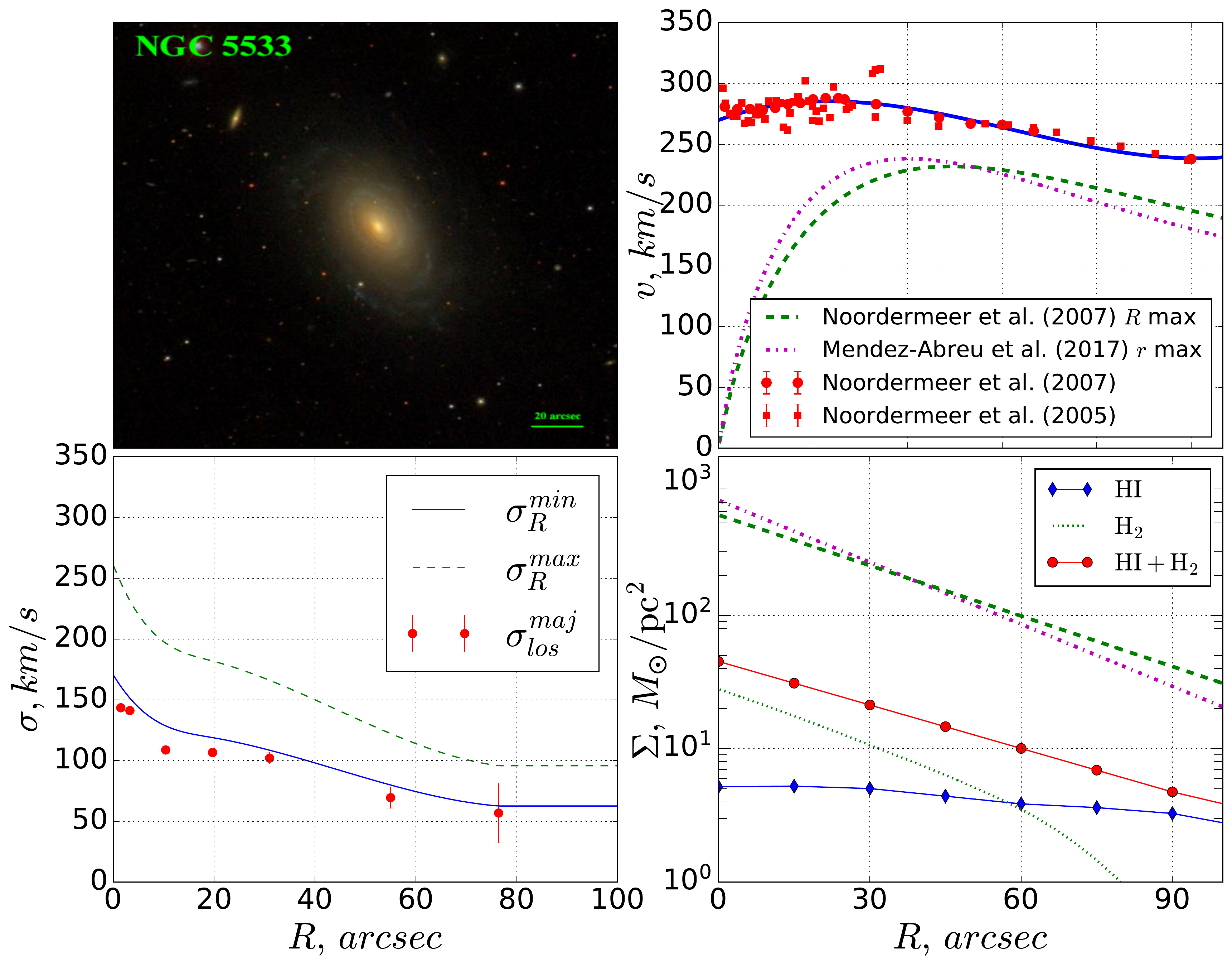}
\caption{The same as in Fig.~\ref{fig:data_338} for NGC~5533. The molecular gas surface densities profile $\Sigma_\mathrm{H_2}$ is shown as a dotted line because its model differ from other galaxies.}
\label{fig:data_5533} 
\end{figure*}
%%%%%%%%%%%%%%%%%%%%%%%%%%%%%%%%%%%%%%%%%%%%%%%%%%%%%%%%%%%%%%%%%%%%%

%%%%%%%%%%%%%%%%%%%%%%%%%%%%%%%%%%%%%%%%%%%%%%%%%%%%%%%%%%%%%%%%%%%%%
${\bf NGC~5533}$ is a bright ($M_R=-22.6^{\mathrm{m}}$) galaxy of Sab type  with a compact bulge. Its inclination angle is around $53\degr$ \citep{Noordermeer_etal2008}. The galaxy possesses one tightly twisted spiral, which shows weak  blue regions of active star formation all along up to $100\arcsec$ from the centre.

The stellar rotation curve and the line-of-sight velocity dispersion profile extend up to $80\arcsec$. The velocity dispersion profile along the major axis $\sigma_\mathrm{los, maj}$ demonstrates the value about 140~km\,s$^{-1}$ in the centre and then decreases almost linearly. The gas rotation curve shows a maximum 280~km\,s$^{-1}$ at about $50\arcsec$ from the centre, after that it slowly decreases, except for the $200\arcsec$ peculiarity. The total amount of atomic hydrogen in the disc is very large and according to \citet{Noordermeer_etal2005} is equal to $3\times10^{10}\, M_{\sun}$. The surface densities $\Sigma_\mathrm{H\,\sc i}$ are exceed 2\,$M_{\sun}$\,pc$^{-2}$ up to $100\arcsec$.

%%%%%%%%%%%%%%%%%%%%%%%%%%%%%%%%%%%%%%%%%%%%%%%%%%%%%%%%%%%%%%%%%%%%%%
\bibliographystyle{mnras}
\bibliography{art}
%%%%%%%%%%%%%%%%%%%%%%%%%%%%%%%%%%%%%%%%%%%%%%%%%%%%%%%%%%%%%%%%%%%%%%
\label{lastpage}
%%%%%%%%%%%%%%%%%%%%%%%%%%%%%%%%%%%%%%%%%%%%%%%%%%%%%%%%%%%%%%%%%%%%%%
\end{document}